# Laser remote magnetometry using mesospheric sodium

Thomas J. Kane[1], Paul D. Hillman[1], Craig A. Denman[1], Michael Hart[2], R. Phillip Scott[2], Michael E. Purucker[3], and Stephen J. Potashnik[4]

## ABSTRACT

We have demonstrated a remote magnetometer based on sodium atoms in the Earth's mesosphere, at a 106-kilometer distance from our instrument. A 1.33-watt laser illuminated the atoms, and the magnetic field was inferred from back-scattered light collected by a telescope with a 1.55-meter-diameter aperture. The measurement sensitivity was 162 nT/√Hz. The value of magnetic field inferred from our measurement is consistent with an estimate based on the Earth's known field shape to within a fraction of a percent. Projected improvements in optics could lead to sensitivity of 20 nT/√Hz, and the use of advanced lasers or a large telescope could approach 1-nT/√Hz sensitivity. All experimental and theoretical sensitivity values are based on a 60° angle between the laser beam axis and the magnetic field vector; at the optimal 90° angle sensitivity would be improved by about a factor of two.

## INTRODUCTION/BACKGROUND

Atoms with an unpaired electron, such as sodium, can have ground states with both angular momentum and a magnetic moment. Such atoms will precess in a magnetic field. The frequency of precession, known as the Larmor frequency, is linearly proportional to the magnitude of the magnetic field. The constant of proportionality is known as the gyromagnetic ratio, and it is known with high precision for sodium and other atoms. Thus a measurement of the precession frequency of an atom determines the magnitude of the magnetic field in which the atom is immersed. Magnetometry based on this kind of measurement using atoms within the instrument is a well-established technology (Budker and Romalis, 2007).

Ninety kilometers above the Earth's surface is a naturally occurring layer of sodium atoms, the residue of meteor ablation, with a density of roughly 4 x $10^9$ atoms per cubic meter, and a thickness of about 10 kilometers (McNeil et al., 1995; Happer, et al., 1994; Plane, 1991). Light at the sodium $D_{2a}$ resonance wavelength of 589.159 nm interacts strongly with these atoms, such that about 4% of the light is resonantly scattered as it passes through the 10-km-thick sodium layer. This scattering from mesospheric sodium is what enables the technology of adaptive optics based on "guidestar" lasers. Laser sources with power up to 50 watts, operating exactly at the sodium resonance wavelength, have been built as guidestar lasers for large telescopes incorporating adaptive optics (Denman et al., 2005).

In 2011, Higbie et. al. proposed a system for remotely measuring the magnitude of the geomagnetic field, typically called *F* by geophysicists, by interrogating sodium atoms in the mesosphere, using a pulsed laser at the sodium resonance wavelength (Higbie et al., 2011). In this measurement, the pulse repetition frequency (PRF) of the laser is adjusted to match the Larmor frequency of the atoms in the magnetic field. If the intensity of the light is in the appropriate range, then the amount of light backscattered by the sodium atoms will be enhanced when the PRF is equal to the Larmor frequency. By tracking the frequency of this resonant enhancement, the magnetic field can be measured remotely.

The existence of a resonance is due to two phenomena. First, the scattering of circularly-polarized photons creates atoms with both angular momentum and a magnetic moment, which precess at the Larmor frequency. Second, these atoms with angular momentum have enhanced backscatter along the axis of the angular momentum. Once precessing, they are optimally oriented for enhanced backscatter once per Larmor cycle time.

Working at the guidestar-laser-equipped Kuiper Telescope (University of Arizona, 2016), one of the facilities of the University of Arizona's Steward Observatory, we have observed this resonance, and have thus measured the magnetic field in the sodium layer. The sensitivity of the magnetic field measurement as characterized by an equivalent noise spectral density was 162 nT/√Hz.

A measurement of magnetic field made by averaging for a period of one hour would have a noise bandwidth of *(3600 sec)*$^{-1}$ *= 0.28 millihertz*. With this bandwidth, the rms uncertainty due to measurement noise would be 2.7 nT.

The following section has four subsections. Subsection 1 derives equations for the sensitivity of the instrument, based on the parameters of the return signal, and assuming that shot noise in the return signal is the predominant noise source. Subsection 2 describes the expected nature of the signal returned from the sodium atoms. Subsection 3 describes our experimental setup and results, and makes comparison with the expected results. Subsection 4 discusses the results, and ways where our system could be utilized or enhanced.

[1]FASORtronics LLC, 1400 Eubank Blvd SE, Albuquerque, NM 87123 USA E-mail: tom.kane@ieee.org, phillman5@gmail.com, craig.a.denman@gmail.com.
[2]University of Arizona, College of Optical Sciences, 1630 E University Blvd, Tucson, AZ 85719.
[3]Planetary Magnetospheres Laboratory, Code 695, Goddard Space Flight Ct, NASA, Greenbelt, MD 20771.
[4]NSWCCD 753/ONR 321MS, 9500 MacArthur Blvd., Bld. 80, Rm. 203, West Bethesda, MD 20817.

Figure 1 is a photograph of the Kuiper Observatory showing the guidestar system in operation, taken from atop Mt. Lemmon, at a 6-km distance. Figure 2 shows, on left, the Kuiper Telescope dome on a moonlit night during operation of the guidestar system and, on right, the 61-inch-diameter (1.55 meters) aperture telescope. Figure 3 shows the "FASOR" laser source, which sits on a 4-foot by 8-foot (1.2-meter by 2.4-meter) optical table.

## DESCRIPTION, RESULTS, AND DISCUSSION

### 1. Instrument Sensitivity

Figure 4 shows a generic resonance shape function $R(f)$. For simplicity, we use a triangle-shaped resonance function; the use of another resonance function would change the sensitivity as expressed by equation 8 by a constant of order one, but would not change the scaling of that expression. The resonance has a full-width at half-maximum (FWHM) of $\Delta f$, in units of hertz. The vertical axis is dimensionless. The resonance sits atop a background level of one. The height of the resonance above the background is $H$, most conveniently thought of in units of percent. The peak of the resonance shape function is at $f = 0\ Hz$.

Just below the frequency of the resonance, in the region of constant positive slope, the slope of the resonance function $dR/df$ is given by

$$dR/df = H / \Delta f. \quad (1)$$

We can use the unitless, zero-centered resonance function of figure 4 to construct a function $S(f,F)$ giving the return signal, in units of photons counted per second, or sec$^{-1}$,[5] as a function of laser pulse repetition frequency $f$ and magnetic field $F$. This function is

$$S(f,F) = S_0\, R(f\text{-}gF) \quad (2)$$

where $S_0$ is the return signal away from the resonance. Note that the resonance is now shifted so that for a magnetic field $F$ the resonance is centered at the Larmor frequency, given by

$$f_L = gF \quad (3)$$

where $F$ is the magnetic field magnitude in nT and $g$ is the gyromagnetic ratio of atomic sodium, equal to 6.99812 Hz/nT, the Landé $g_f$ factor times the Bohr Magnetron, divided by Planck's constant (Auzinsh et al., 2010).

If pulse repetition frequency $f$ were held fixed on the positive-sloped side of the resonance, and if there were no measurement noise, then any change in $S$ would imply that $F$ had changed. The ratio of the inferred change in $F$ to the observed change in $S$ is the calibration factor of the magnetometer, $C$, given by

$$C \equiv (\partial S/\partial F)^{-1} = -\Delta f / (gS_0H). \quad (4)$$

Measurement noise will create changes in $S$ that limit the ability to measure small changes in $F$. In the following, we find the magnitude of the change in $F$ that is needed to make a change in $S$ equal to the change in $S$ that is due to noise.

The statistical variance $<\Delta S^2>$ in the count rate of a random Poisson process with average count rate $S_0$ is

$$<\Delta S^2> = 2S_0 BW \quad (5)$$

where $BW$ is the bandwidth of the counting process, in hertz (Engleberg, 2007). This is "shot noise," and it creates an apparent variance in $F$, which we call $<\Delta F^2_{SN}>$, that puts a lower limit on the ability of the magnetometer to measure an actual change in $F$. This is called the shot-noise-limited sensitivity of the instrument.

[5] *It will be convenient to use the unit msec$^{-1}$ instead sec$^{-1}$, Of course 1 msec$^{-1}$ = 1000 sec$^{-1}$.*

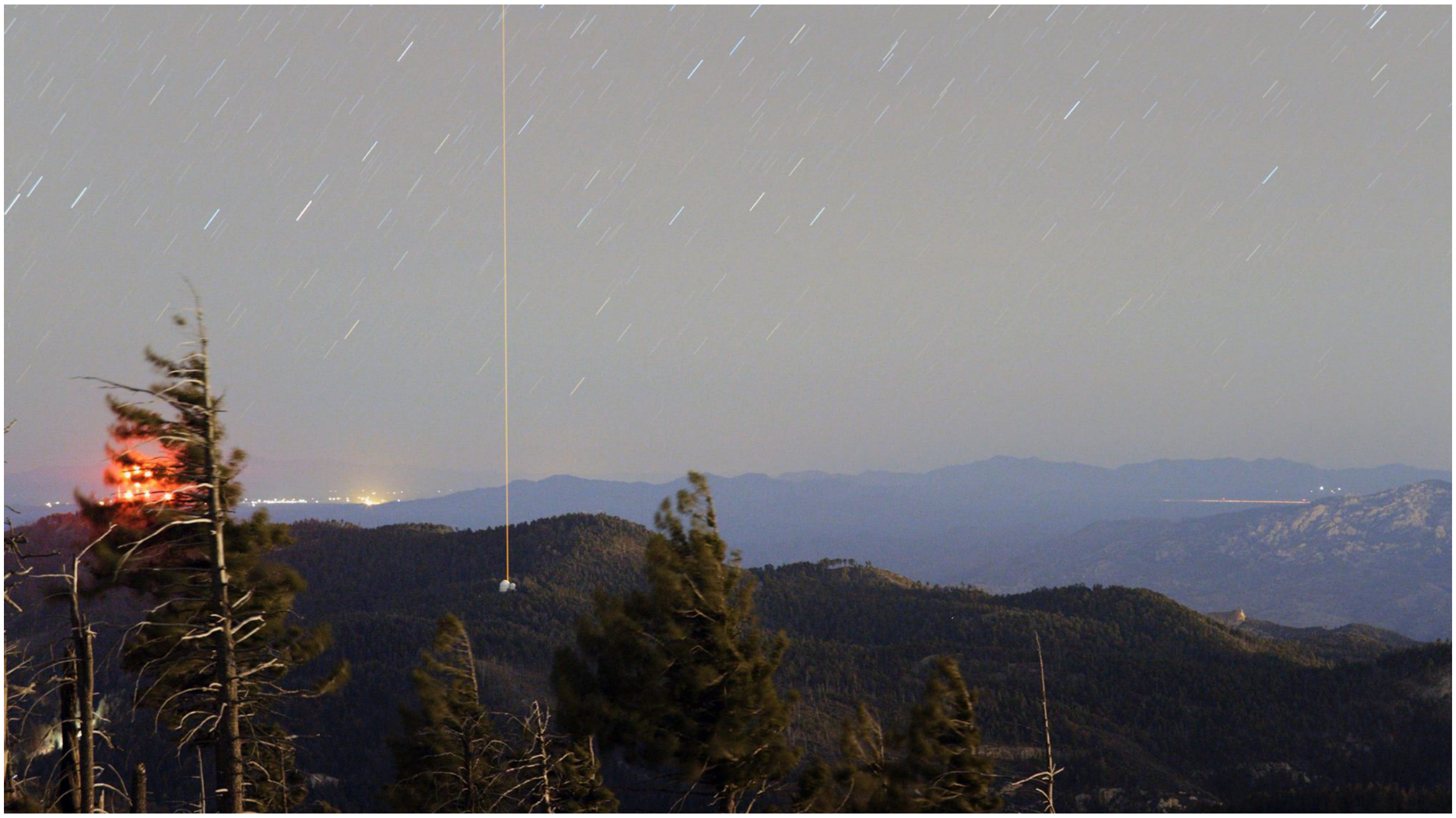

Figure 1. Kuiper Observatory, photographed from atop Mt. Lemmon, with guidestar in operation. On this night, beam was pointed straight up, while for magnetometry the beam launch angle was 60°.

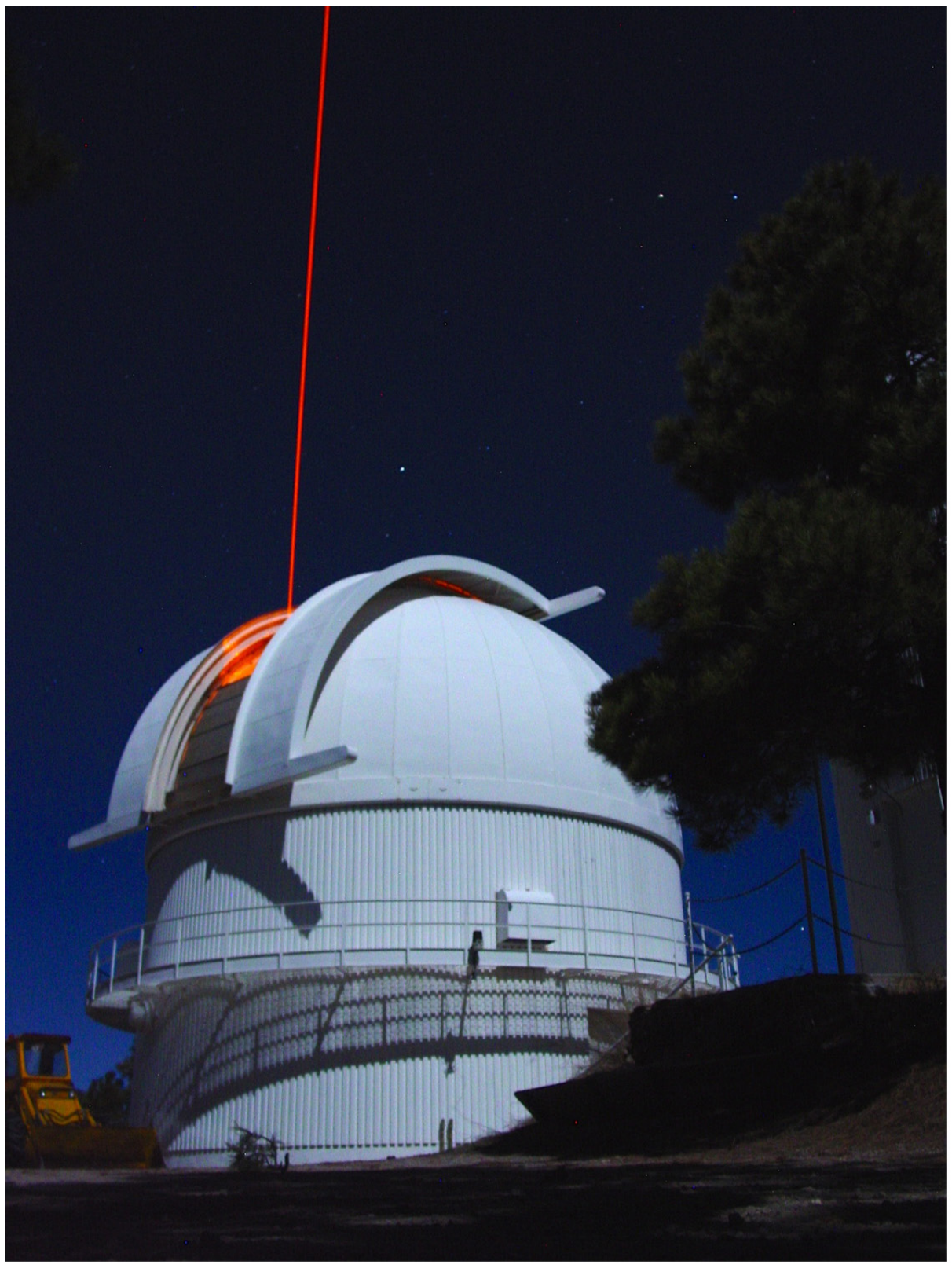

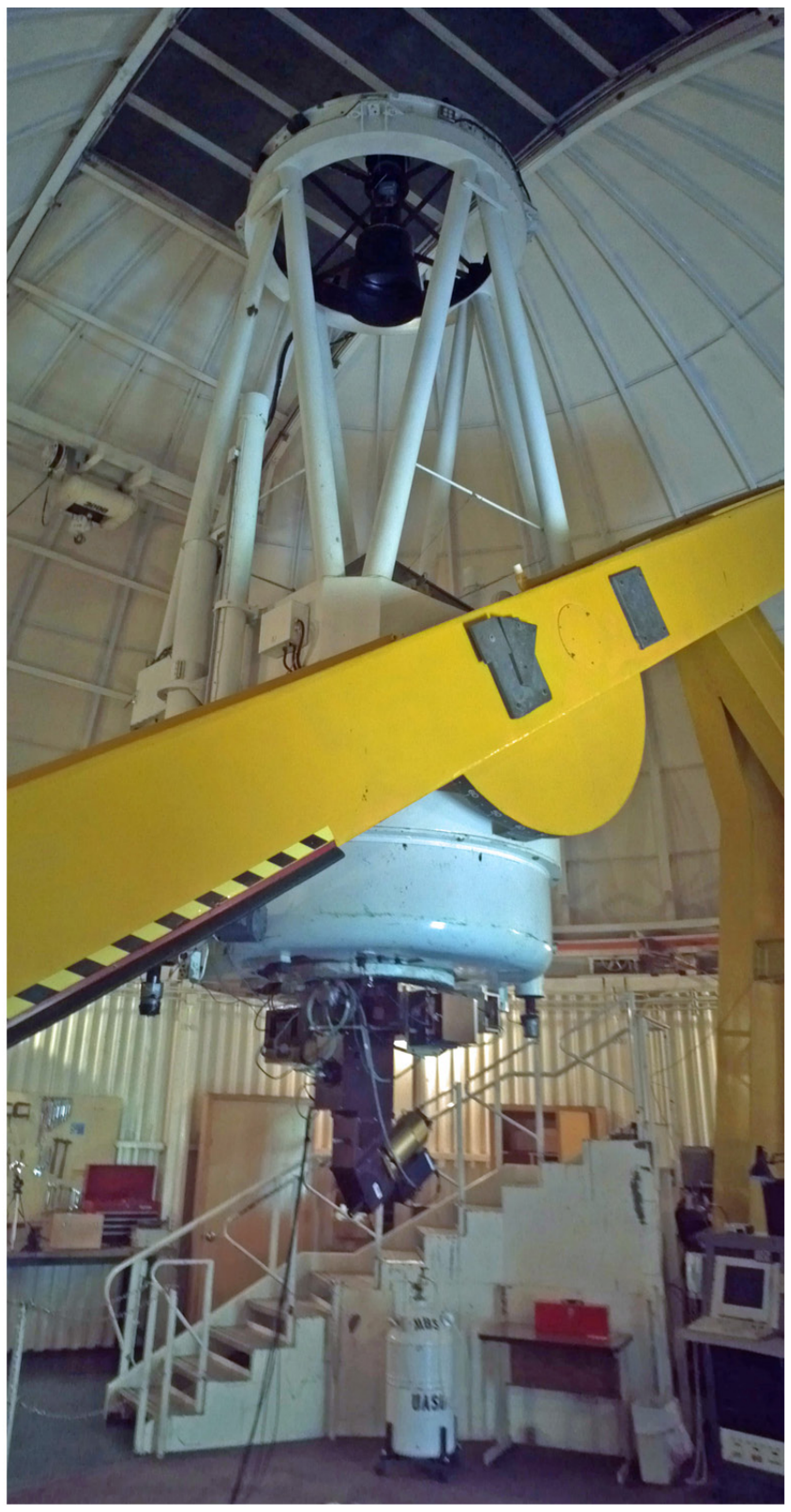

Figure 2. Left, the Kuiper Telescope dome on a moonlit night, with guidestar in operation. Inside the dome, on the right, the Kuiper Telescope used to receive the signal from the sodium layer.

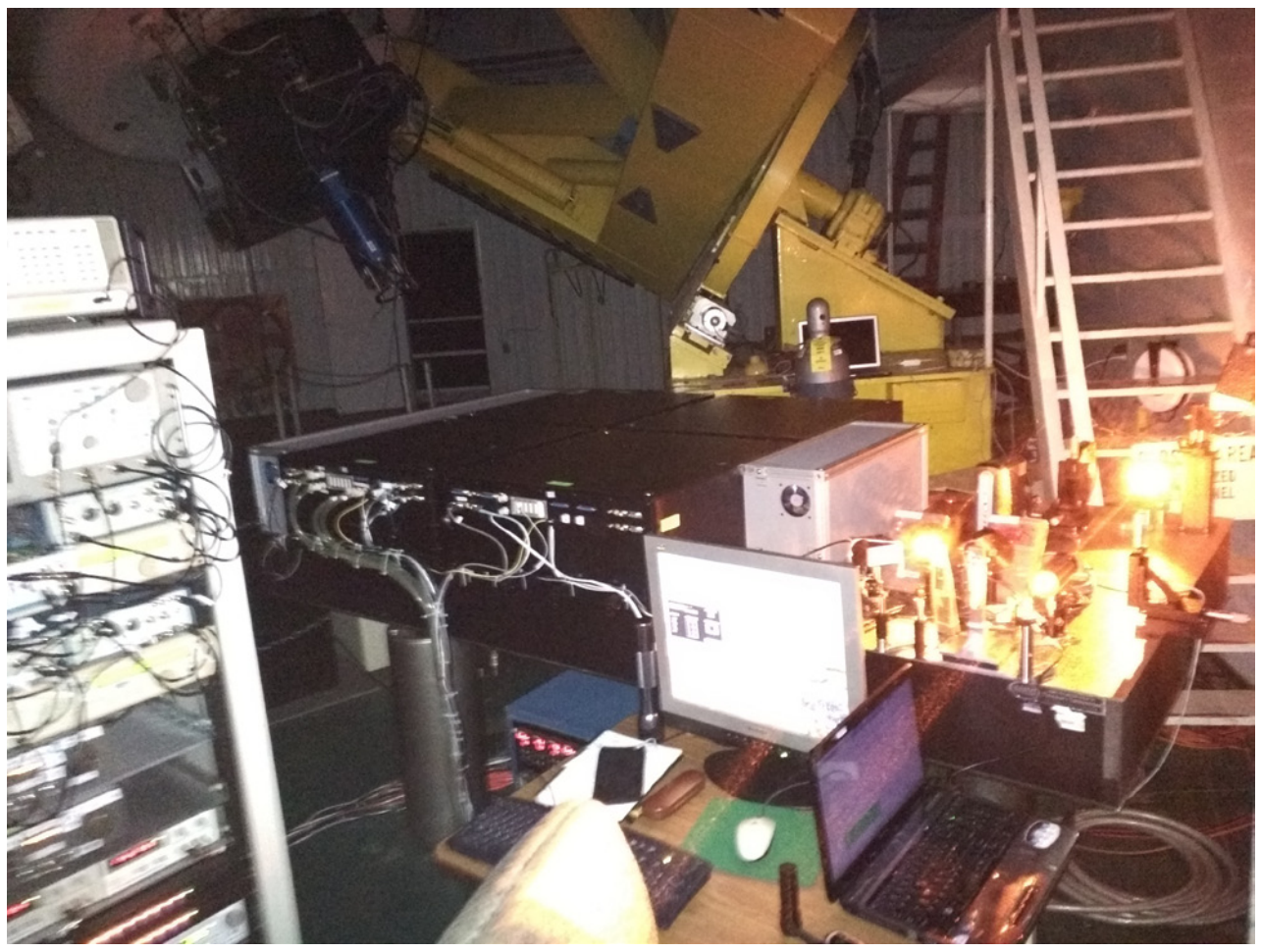

Figure 3. The "FASOR" laser in operation in the telescope dome. Four boxes contain the FASOR components. One is a single frequency 1319-nm Nd:YAG laser, and another is a similar laser at 1064 nm. The third box contains a resonant frequency summing system. The final box contains a wavemeter and a sodium frequency reference, and the acousto-optic modulator. Components in the open on the table include the polarization-setting waveplates and the launch telescope. Control electronics are in the rack on the left and in the gray boxes on the optics table near the laser boxes.

The shot-noise-limited sensitivity $<\Delta F^2_{SN}>$ of the instrument is given by

$$<\Delta F^2_{SN}> \; = C^2 <\Delta S^2> \tag{6}$$

$$= 2\,\Delta f^2 BW\,/\,(g^2\,S_0\,H^2)\,. \tag{7}$$

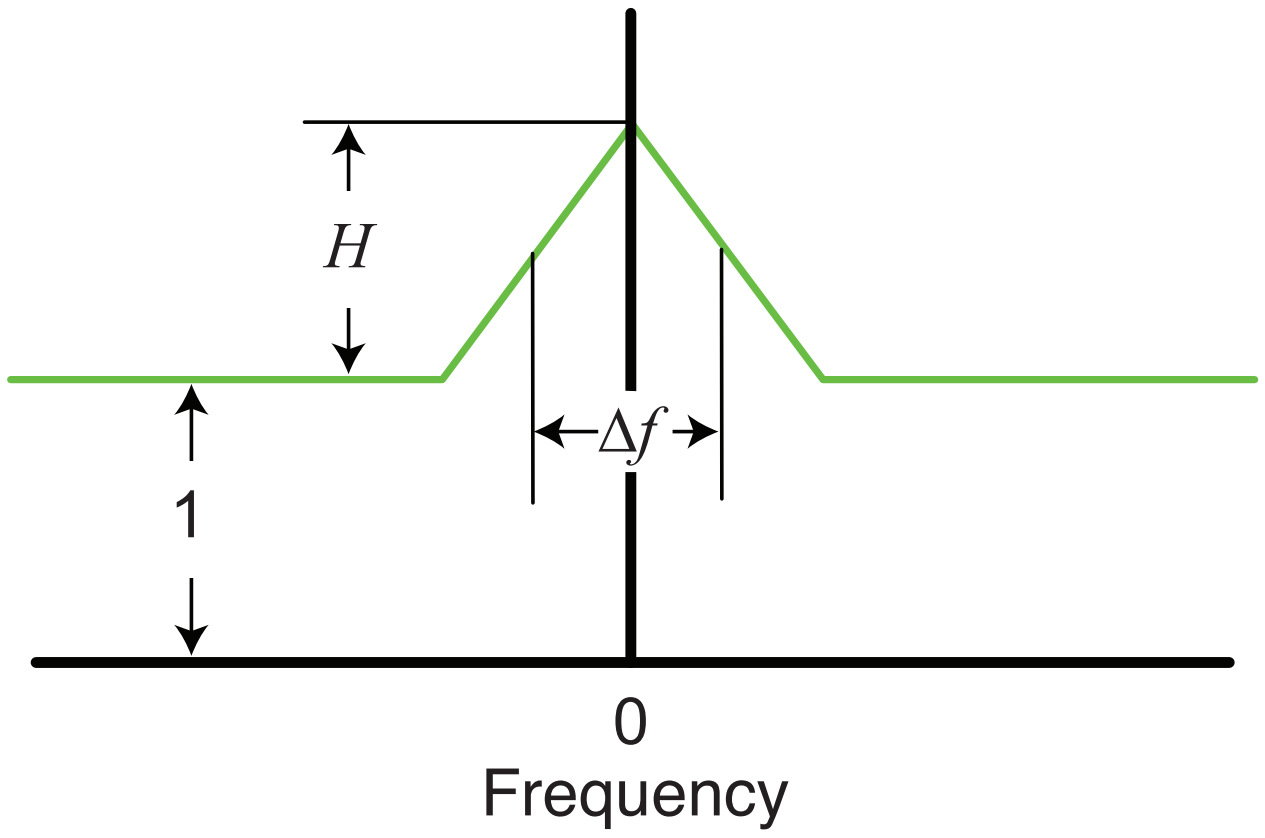

Figure 4. Triangle resonance shape, with key parameters labeled.

The first equivalence is the result of the fact that by definition a change in signal $\Delta S$ creates an apparent change in magnetic field of $C\Delta S$, and the second equivalence is the result of substituting equations 4 and 5 into 6. A simplifying assumption used here is that $H<<1$, so that $S \approx S_0$, and the two are considered equal for purposes of noise calculation.

The value $<\Delta F^2_{SN}>$ is a variance, and has units of (nT)$^2$. A commonly used related value is the square root of the power spectral density of the noise-driven apparent change in $F$, which has units of nT/√Hz, and is called the *noise-equivalent signal, NES*. This value is given by

$$NES \equiv \sqrt{\frac{<\Delta F_{SN}^2>}{BW}} = \frac{\sqrt{2}\Delta f}{gH\sqrt{S_0}} \ . \qquad (8)$$

If the resonance function $R(f)$ is a Lorentzian, and if $f$ is selected to be at the steepest point on the resonance (that is, the inflection point), then a factor of $4\sqrt{3}/9 \approx 0.77$ multiplies the right side of the equation. If it is not the case that $H<<1$, the analysis is somewhat more complicated, since the shot noise would then depend on where you are on the resonance. But for analyzing the magnetometer described in this paper, equation 8 is adequate.

A small value of $NES$ is desirable. Thus we want a large value of signal $S_0$, which can be accomplished with a more powerful laser or a larger collection telescope. We also want the smallest possible value of $\Delta f$, the full-width at half-maximum resonance, and a large value for the dimensionless resonance strength, $H$. Though these parameters are not as readily changed, they do depend on the intensity of the laser at the sodium layer, and on other laser properties. The gyromagnetic ratio $g$ is of course a constant.

Equation 8 applies regardless of whether the signal $S_0$ is entirely from sodium signal return, or if it is due to some other source, as long as shot noise is the dominant noise. If there is light from a signal other than sodium, that light will not show a resonance, and $H$ will be reduced, more than offsetting the benefit of the increase in $S_0$ in equation 8.

## 2. Expected Signal

The expected value of the signal $S_0$ is given by the equation

$$S_0 = \Psi\ A\ P\ N\ \eta_{atmosphere}{}^2\ \eta_{optics}\ \eta_{detector}\ \sin(\theta)\ /\ z^2 \qquad (9)$$

with the following definitions:

$\Psi$ – the sodium backscatter cross section, in units of (photons/second/steradian/atom)/(W/m$^2$);[6]
$A$ – the collection area of the telescope, in units of m$^2$;
$P$ – the time-averaged power transmitted to the sky, in units of watts;
$N$ – the sodium atom density per unit area looking upward, in units of atoms/m$^2$;
$\eta_{atmosphere}$ – the one-way efficiency of atmospheric transmission;
$\eta_{optics}$ – the efficiency of the telescope optical train;
$\eta_{detector}$ – the quantum efficiency of the detector, in units of counts per incident photon;
$\theta$ – the elevation angle of the telescope, with 90° being straight up, and;
$z$ – the vertical distance to the sodium layer in units of meters.

Equation 9 is essentially the LIDAR equation (Wandinger, 2005). When the laser is at an angle other than 90°, the distance to and the thickness of the sodium layer are both increased by $1/\sin(\theta)$. Therefore, the value $z$ in equation 9 is the vertical component of the distance to the scatterers, not the total distance.

Table 1 gives estimated values for each parameter of equation 9, and a resultant value for $S_0$, for the experimentally realized case that will be described in Section 3.

The first value of $S_0$ corresponds to expected return when the laser is operated continuously. The second is what is expected when it is pulsed with a 35% duty cycle, with the time between pulses much shorter than the time spent passing through the sodium layer. In this latter case, the average power in the sodium layer is 35% of the cw output of the laser, but the return signal is continuous since many pulses are in the sodium layer at once.

With $S_0$ estimated, we still need an estimate of resonance height $H$ and resonance width $\Delta f$ in order to complete the right side of equation 8 and have an estimate of the noise-equivalent signal in units of nT/√Hz.

In order to calculate $H$ and $\Delta f$, we modeled the sodium physics using code written by Simon Rochester, based on the open-source LGSBloch Mathematica package, which is an extension of the Atomic Density Matrix package (Rochester, 2007). The following parameters are the inputs to the model:

- *Laser wavelength.* We assumed a laser centered at the $D_{2a}$ line of sodium, at 589.159 nm.
- *Laser linewidth and sidebands.* We assumed essentially zero linewidth, with no sidebands. These assumptions are well met for the Frequency Addition Source of coherent Optical Radiation (FASOR) we have used.
- *Laser intensity at the sodium atoms.* We varied this parameter over our range of interest.
- *Laser modulation.* We assumed that the laser was pulsed, with square pulses at a constant pulse repetition frequency (PRF) and duty cycle. From model run-to-run, we varied PRF and duty cycle. Though our modelling suggested a PRF closer to 20% duty cycle might be best (Denman et al., 2012), we found from our on-sky measurements that a duty cycle of 35% (pulse length equal to 35% of the PRF cycle) was close to optimal, and for simplicity all results shown in this paper are for a 35% duty cycle.
- *Laser polarization.* All results are for circular polarization, which is optimal.
- *Magnetic field.* We assumed 45.3 μT, from NOAA-maintained NGDC software “Geomag” version 7.0 using the IAGA geomagnetic model coefficients predicted for 92-km altitude over Tucson, Arizona, United States (Thébault, E., 2015).
- *Magnetic field orientation.* The optimal orientation for creating a strong Larmor resonance is to propagate the

**Table 1. Parameters allowing an estimate of return signal.**

| | |
|---|---|
| $\Psi$ | 240 photons/sec/sr/atom/(W/m$^2$) |
| $A$ | 1.8 m$^2$ |
| $P$ | 3.8 watts |
| $N$ | 4 x 10$^{13}$ atoms/m$^2$ |
| $\eta_{atmosphere}$ | 84% |
| $\eta_{optics}$ | 70% |
| $\eta_{detector}$ | 0.27 counts/photon |
| $\theta$ | 60° |
| $z$ | 92 kilometers |
| $S_0$, using equation 9 | 896 msec$^{-1}$ |
| $S_0$, 35% duty cycle | 314 msec$^{-1}$ |

[6] *The backscatter cross-section can also be expressed in units of area, according to the equation $\sigma = \Psi\ h\nu$, where $h\nu$ is the photon energy, 3.4 x 10$^{-19}$ J. This yields a value of $\sigma$ = 8.2 x 10$^{-17}$ m$^2$, an extraordinarily large value, indicating the extraordinary ability of sodium to backscatter light at its resonance wavelength. Equation 9 would then be changed by replacing P with P/hν.*

laser, such that at the 92-km altitude of the sodium layer, the direction of propagation of the laser is perpendicular to the Earth's magnetic field. Ideally, this direction would have been 9.75° azimuth and 31.03° elevation. However, due to telescope-mount constraints and the restraints placed on us by the Federal Aviation Administration (FAA), we were limited experimentally to 25.1° azimuth and 59.7° elevation. Thus, the angle between the geomagnetic field and the laser propagation was close to 60° and that is what we modeled. Our earlier report (Denman et al., 2012), describes how the magnetometer sensitivity becomes worse as this angle decreases from 90°; thus, our sensitivity at this angle is about 2x worse than if we had propagated optimally.

All theoretical results in this paper use this angle of 60°. The best-case angle of 90° results would be twice better. But we were limited to a 60° angle at the Kuiper Telescope, so we consistently use that value. The optimal 90° angle may likewise only be possible in the tropics; 60° or better should be possible over most mid-latitude locations.

- *Atom environment.* We used mean times between interatomic collisions and spin relaxation appropriate for the high mesosphere.

Holzlöhner et al. (2010) used almost the same model of sodium atomic physics for their paper on expected guidestar return, and that paper is a good reference if further understanding of sodium modeling is desired.

Figures 5 and 6 show results of model calculations of the sodium resonance width $\Delta f$, in units of kilohertz, and the resonance height $H_{Na}$, in units of percent, as a function of the time-averaged intensity of light at the sodium atoms. All atoms are assumed to see the same intensity, as would be the case with an ideal "top-hat" beam shape. (Here it is assumed that all of the signal $S_0$ is due to sodium, and for this case there is no distinction between $H$ and $H_{Na}$. Later we will define the difference.)

The atmospheric pressure is an important variable. At lower pressure, the mean time between collisions of sodium atoms with air molecules is longer, and the light has more time to act on the sodium atoms before their angular momentum states are randomized by collisions, so the same degree of synchronized angular momentum can be achieved at lower light intensity. Over the ~10-kilometer thickness of the sodium layer, atmospheric pressure varies by a factor of six. Figures 5 and 6 each show three cases: for atoms at the center of the sodium layer at 92 km, and for atoms above 25% and 75% of the sodium layer, at 89 km and 95 km. The atmospheric pressure at these three altitudes is 0.14, 0.23, and 0.084 Pa. The parameters in the model corresponding to atmospheric pressure are the rates of velocity changing collisions and spin relaxation, which were assumed to be linear functions of pressure.

It can be seen from inspecting figures 5 and 6, in light of equation 8, that there is an optimum intensity for sensitive magnetometry. To minimize the sensitivity, as is desired, the ratio $\Delta f/H$ must be minimized. At very low intensity, $H$ approaches zero, and $\Delta f$ is at a minimum but non-zero value, so $\Delta f/H$ goes to infinity. At higher intensity, $H$ rises steeply at first and saturates, while $\Delta f$ rises linearly, so $\Delta f/H$ will also rise linearly. The minimum $\Delta f/H$ ratio clearly occurs at an intensity somewhat below 1 W/m$^2$. Using equation 8, with $g = 6.99812$ Hz/nT, taking $S_0$ for 35% duty cycle from table 1, and taking $H$ and $\Delta f$ from figures 5 and 6 at the altitude of 92 km, we calculate a sensitivity of 19 nT/√Hz at 0.35 W/m$^2$ and 21 nT/√Hz at 1.05 W/m$^2$, and larger values at both 0.1 W/m$^2$ and 3.5 W/m$^2$. It will be seen that experimentally observed values are almost an order of magnitude worse, primarily because values of both $S_0$ and $H$ turned out lower than expected.

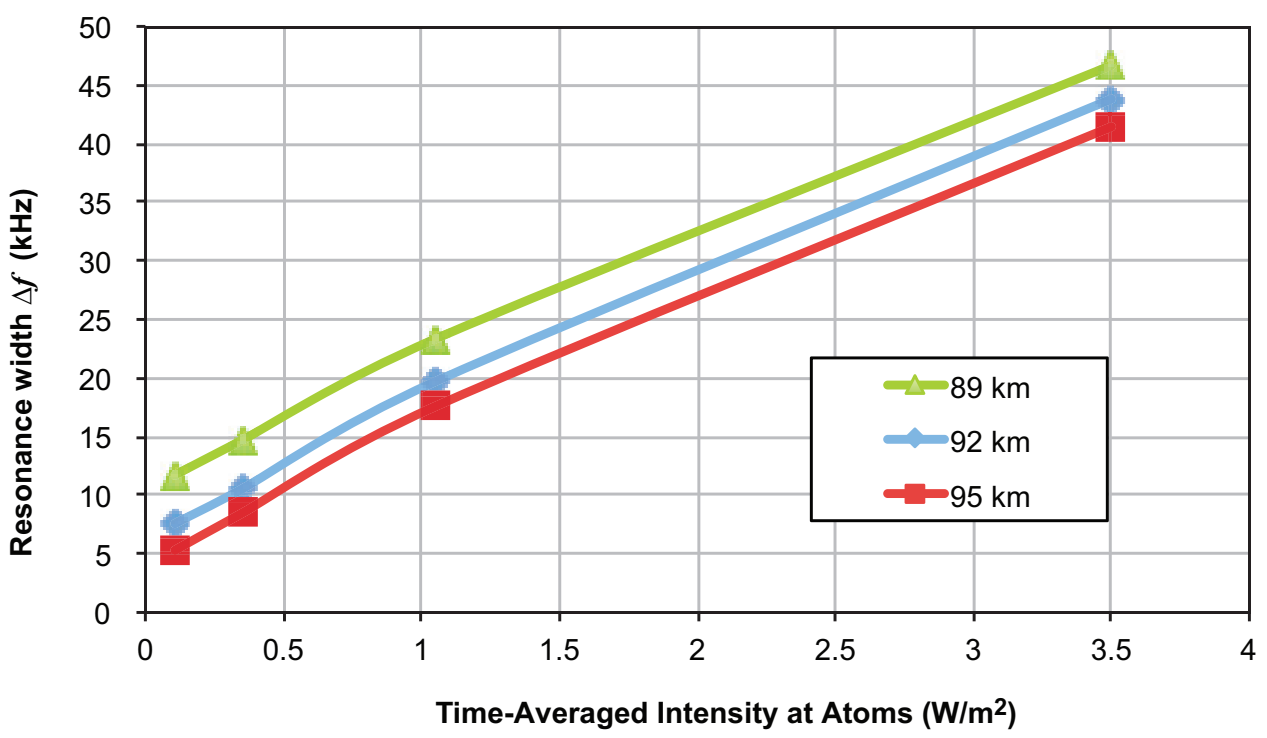

Figure 5. Model calculation of resonance width $\Delta f$ as function of the (time-averaged) intensity of light at the atoms, for three different altitudes covering the range of the sodium layer.

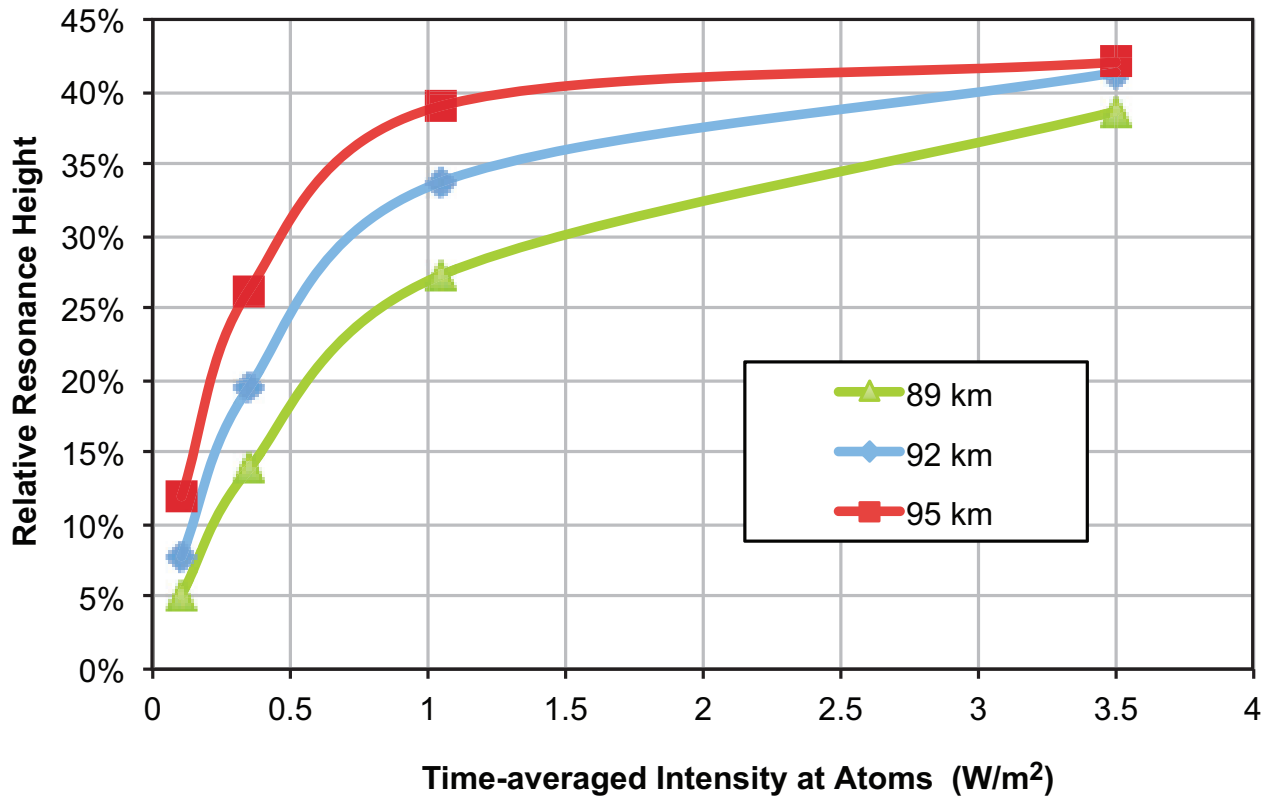

Figure 6. Model calculation of resonance height $H$ as function of the (time-averaged) intensity of light at the atoms, for three different altitudes covering the range of the sodium layer.

## 3. Experimental System

Figure 7 diagrams the experimental system that we built.

*Transmitter*

A continuous wave, single-frequency FASOR provides the coherent light. This source generates light at the wavelength of 589.159 nm by frequency summing the output of two single-frequency lasers, one at 1319 nm and the other at 1064 nm. (All available coherent sources at 589 nm are based on nonlinear conversion from infrared lasers, since no practical laser at 589 nm is currently available. Dye lasers at 589 nm have been built but have many problems.) We call our light source a Frequency Addition Source of coherent Optical Radiation, or FASOR. The output of the FASOR is up to 10 watts of diffraction limited, single-frequency light, easily tunable to exactly the sodium $D_{2a}$ resonance wavelength of 589.159 nm. This FASOR was designed and built originally for use as a concept prototype guidestar laser (Denman et al., 2004; Bienfang et al., 2003). It now belongs to the Steward Observatory and is installed in the Kuiper Telescope dome, atop Mount Bigelow, in the Catalina mountains north of Tucson, Arizona.

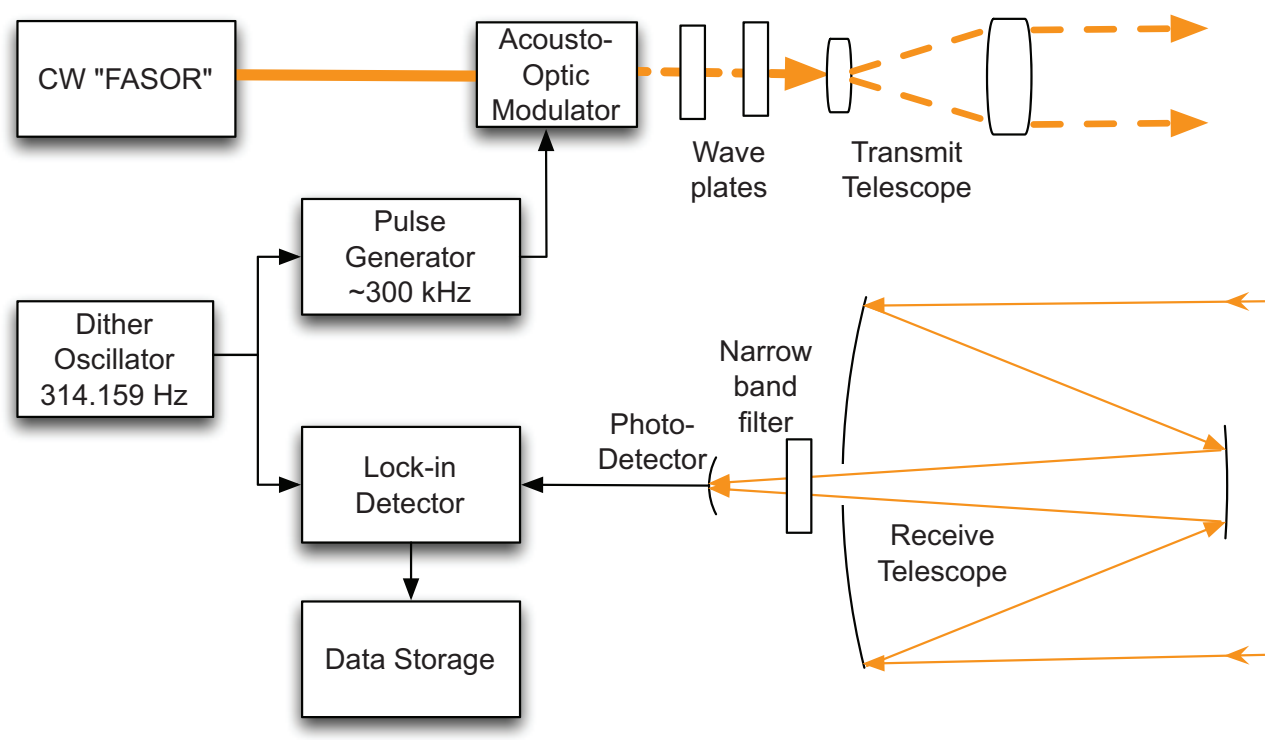

Figure 7. Components of the magnetometer.

This light is passed through an acousto-optic modulator, model 35085-3 from Gooch & Housego. This device can deflect up to 80% of the light, with a rise and fall time of 300 nanoseconds. The deflected, first-order output beam from the modulator was utilized so that the off-state power was zero watts. After modulation, the polarization state of the light could be adjusted into any arbitrary state by a quarter waveplate and a half waveplate, in combination. A beam expanding telescope expands the slightly elliptical beam so that it has $1/e^2$ radii along the two transverse axes of 38 x 26 millimeters. Not shown in the diagram is a flat mirror that directs the beam to the sky. We used a shear plate to optimize the telescope for collimation of the light. We measured both the power and the polarization state of the light after the final flat mirror. The polarization was adjusted to circular. The total power exiting to the sky was 3.8 watts, when the acousto-optic modulator was continuously deflecting the beam.

*Receiver*

We used the 61-inch (1.55 meter) Kuiper Telescope as the receive telescope to observe the beam in the sodium layer. This telescope has an effective aperture of 1.8 $m^2$, taking into account the light blocked by the 0.41-meter-diameter secondary mirror.

The photodetector was a Hamamatsu Multi-Pixel Photon Counter (MPPC), model 12662-150. (Other vendors call this device a silicon photomultiplier or an avalanche photodiode array.) This module contains a semiconductor detector in the form of a 1-mm square. The quantum efficiency is specified as 27%. The dark signal is specified to be below 40 femtowatts, equivalent to a dark count rate of 32 $msec^{-1}$ at 589 nm. The internal avalanche gain of this device is large enough that the detection system is limited by the shot noise in the dark signal, not by noise in the electronics.

A narrow-band optical filter, a Chroma model ET592/21x, was used to eliminate light that was not near the sodium wavelength. It had a pass band of 21 nm centered at 592 nm, and 97% transmission at 589 nm. Without this filter, signal due to background light was significant; with the filter in place, we could operate under a full moon with no significant signal from background light.

*Signal Acquisition System*

In an idealized system, free of any noise other than shot noise in the detected signal, we could map out the Larmor resonance by slowly scanning the pulse repetition frequency of the transmitted light. Once mapped out, we could then tune the PRF to the point of steepest slope on the side of the resonance. Changes in signal detected would then correspond to changes in magnetic field $F$, as per equation 4.

In reality, changes in return signal due to scintillation, wind turbulence, the atmospheric effect that causes stars to twinkle (Dravins et al., 1997; Ryan et al., 1997), laser power fluctuations, beam pointing fluctuations, and other low-frequency disturbances lead to changes in return signal significant compared to the resonance height $H$. The technique of phase-sensitive or lock-in amplifier detection allows the measurement to be moved to a higher frequency, where shot noise is the dominant noise. The lock-in amplifier we used was model 7230 from AMETEK Signal Recovery.

To implement phase-sensitive detection, the PRF of the system is dithered. This allows the PRF to be moved quickly on and off the resonance, or to be moved quickly from one side of the resonance to the other. The lock-in detector looks for a signal at the dither frequency, but with a phase that takes into account the delay due to the ~200-km round trip of the light.

In practice, a dither frequency of a few hundred hertz was adequate to achieve shot-noise-limited detection (Denman et al., 2012). We used a dither frequency of 314.159 Hz. The frequency synthesizer used was the Rigol model DG1062Z, which had all of the features needed to create the frequency-dithered pulse train with a few parts per billion frequency stability, while also providing a reference signal at the dither frequency for the lock-in detector. Thus both the "dither oscillator" and the "pulse generator" of figure 7 are included in the Rigol device. The accuracy of the magnetic field measurement will be directly limited by the accuracy of the frequency synthesizer.

*Characterizing the beam in the mesosphere*

Figure 8 shows the beam in the mesosphere, as recorded by a CCD camera in the same location as that used by the Hamamatsu MPPC. The wedge of light coming in from the top is light scattered from air due to Rayleigh scattering. The spot centered in the box is the sodium signal. The separation between the sodium return and the Rayleigh return is due to the fact that the axis of the transmit telescope is offset by about two meters from the axis of the receive telescope, so that the high-density region of the atmosphere where Rayleigh scattering is significant is seen offset from the sodium spot. The transmit axis is launched above the receive telescope, as is seen in the image in figure 8.

The box in figure 8 corresponds to the size of our 1-millimeter-square detector at an auxiliary focal plane of the telescope. The "plate scale" at the focal plane is 20 arcseconds per millimeter, or 97 microradians per millimeter. At a distance of 106 km, the side of the box corresponds to *106 km x 97 µrad = 10.3 meters*. A fit of the beam in the box gives a 1/e$^2$ radius of 2.6 meters. Since atmospheric effects will increase the apparent size above the actual size, this is a good upper limit on our estimate of the beam size in the sodium layer.

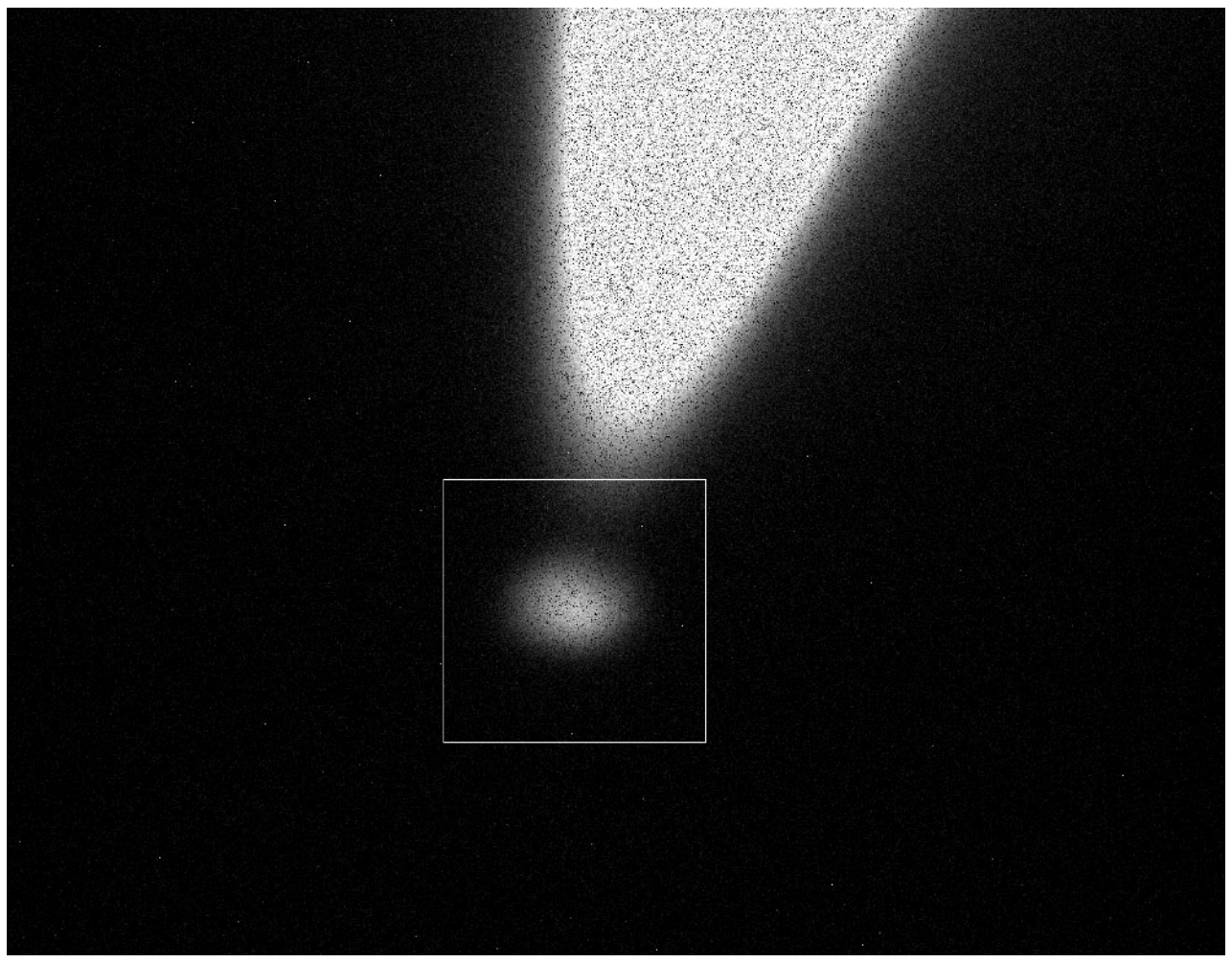

Figure 8. Image of sodium signal (in box) and non-sodium "Rayleigh scatter" signal, above. The box corresponds to our 1-mm-square detector. Its image at a range of 106 km is a 10.3-meter square. The black dots, or image graininess, are an artifact of how we correct for "hot" pixels in our camera. (All pixels that exceed a threshold level in a dark exposure were set to black in corrected exposures.)

A lower limit on beam size can be calculated by assuming that the propagation from the transmit telescope to the mesosphere at 106 km is diffraction limited, so that propagation is according to standard Gaussian beam equations. Starting at 38 x 26 millimeters, this leads to a spot at the 106-km range with 1/e$^2$ radius of 0.52 x 0.76 meters.

These limits on beam size allow us to calculate beam intensity in the mesosphere. Assuming 3.8 watts launched, and 84% atmospheric transmission, the intensity on the beam axis, given by $2\ P\ \eta_{atmosphere} / (\pi\ w_x w_y)$ where $w_x$ and $w_y$ are the beam radii in the two dimensions and other variables are as in equation 9, must be in the range from 0.3 to 5.1 watt/m$^2$. Note that this is the spatial and temporal peak intensity – not the time averaged, or spatially-averaged, intensity. Our duty cycle when collecting signal for magnetometry is 35%, so time-averaged intensity, on beam axis, is 35% of what we have just stated – that is, between 0.1 and 1.8 watt/m$^2$. This can be compared to the range from 0.35 watt/m$^2$ to 1.05 watt/m$^2$ where optimum sensitivity is expected, according to figures 5 and 6.

### *Identifying the sodium return signal*

On March 25, 2016 we gathered return signal data. The first data set, shown in figure 9, was done at the low pulse repetition frequency of 700 Hz, with pulses of duration 0.29 msec. At this low PRF, the signal from the sodium layer can be separated from closer sources by time of flight. (At the higher PRF we used for magnetometry, this is not possible, since the distance between adjacent pulses is short compared to the thickness of the sodium layer, so that multiple pulses are in the sodium layer simultaneously, and the return signal is continuous.) Figure 9 shows return data as a function of time after the beginning of pulse transmission, averaged over multiple cycles to improve signal-to-noise.

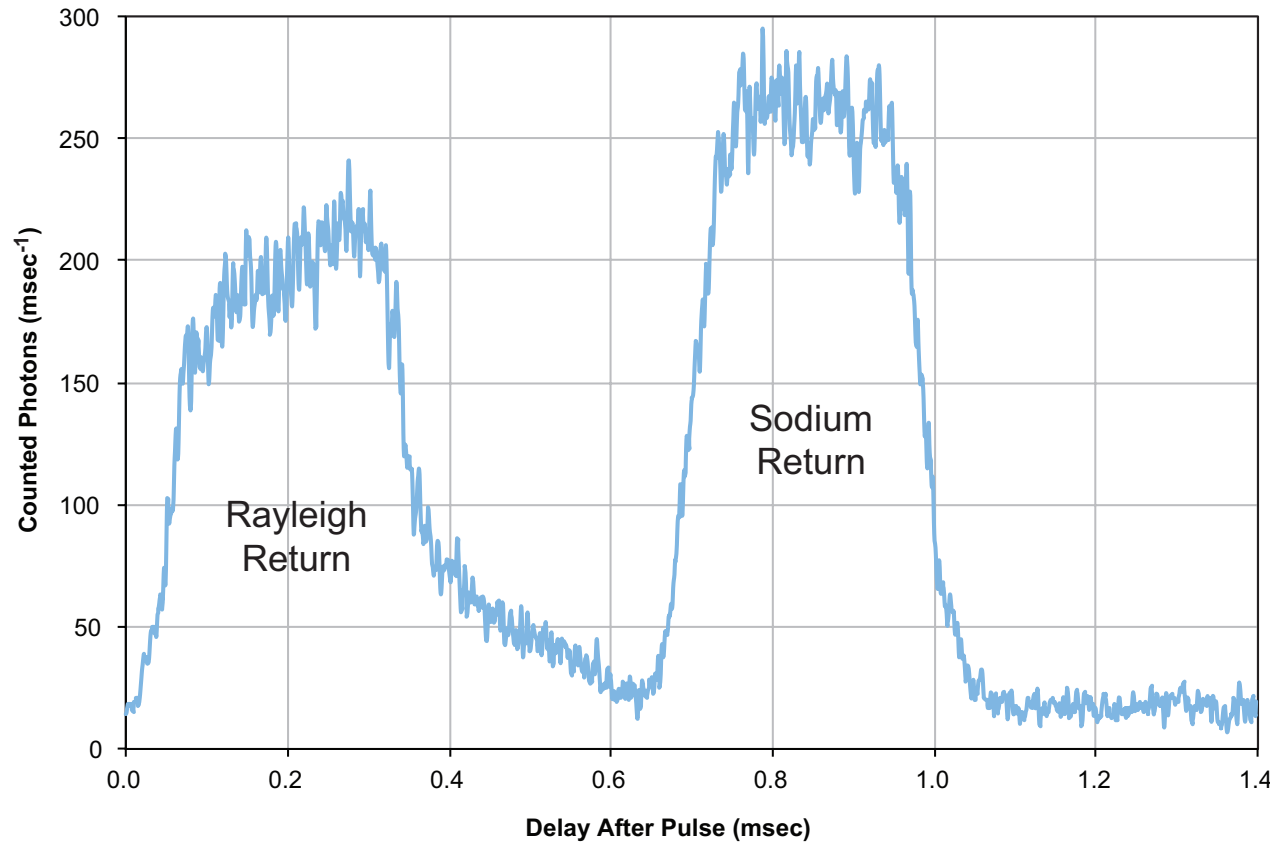

Figure 9. Transmitted light was pulsed at 700 Hz, with 0.29 msec pulses. Return is plotted as a function of time after the beginning of the transmitted pulse. The first received pulse is from non-sodium sources. The pulse from 0.65 msec to 1.05 msec is the sodium signal, which is delayed by the roughly 0.7 msec time of flight. Nearly 53% of the integrated signal is due to sodium. Data was averaged 571 times to improve signal-to-noise.

The signal before 0.6 msec is not due to sodium but, instead, is Rayleigh atmospheric scattering. Given that the image of figure 8 shows good separation between sodium and Rayleigh, i.e. little Rayleigh in the drawn box, we were surprised by the significant amount of non-sodium light detected. The ratio of sodium signal to total signal (ignoring dark signal), as found by integrating the data of figure 9, is 53%.

The sodium signal centered at 0.85 msec has a level, after dark count subtraction, of 250 msec$^{-1}$. This is disappointing compared with the predicted value from table 1 of 896 msec$^{-1}$. The low level of signal may be due to a lower than expected value of sodium atom density *N*, which can vary by a factor of two. It is also likely that there are unexpected losses in the telescope optics that reduce $\eta_{optics}$ below the 70% value of table 1. The signal after 1.1 msec is dark signal. The dark count is estimated from the data of figure 9 to be 17 msec$^{-1}$, comfortably below the 32 msec$^{-1}$ derived from the detector specification.

### *Measurement of Larmor resonance*

The data set of figure 10 shows the magnetic resonance. For this data set, the PRF was dithered by ±15 kHz from its average frequency $f_{avg}$. Thus the PRF is never at this average; it is either 15 kHz above or 15 kHz below. For over 90 minutes, we swept $f_{avg}$ from 283 kHz to 372 kHz. The dither frequency – the frequency at which the PRF makes a full cycle from $f_{avg}$ *+ 15 kHz* to $f_{avg}$ *- 15 kHz* and back – was 314.159 Hz.

The signals plotted in figure 10 are the modulation seen on the return, at the dither frequency, in phase and 90° out of phase (in-quadrature) from the delayed dither frequency. We adjusted the lock-in phase delay to compensate for the time of flight, with the goal of putting all signal into the in-phase channel. The phase correction used was 82.7°, which at 314.159 Hz is equal to *(82.7° /360°) / 314.159 Hz = 0.731 msec*, or a round-trip distance of 219 km. For our telescope angle of 59.7°, and site elevation of 2.5 km, this means phase was optimized for atoms at an altitude of *219 km x sin(59.7°) / 2 + 2.5 km = 97.1 km*.

When the return signal is larger during the first half of the dither cycle, corresponding to the time when light pulsed at $f_{avg}$ *+ 15 kHz* is returning to the telescope, the in-phase value from the lock-in is positive; conversely, when pulses with PRF $f_{avg}$ *- 15 kHz* create an increase in return, the in-phase signal is negative. Thus the positive signal peaks 15 kHz below the Larmor frequency, and its negative image appears 15 kHz above the Larmor frequency. Thus the data of figure 10 shows the underlying resonance, but convolved with a function that creates one peak shifted -15 kHz, and another peak inverted and shifted +15 kHz.

If lock-in phase were adjusted ideally, the in-quadrature signal should be pure noise. However, since the signal has a range of delay times, there is no single phase value that compensates for all ranges; both the nearer and farther sodium atoms contribute some signal to the in-quadrature channel. The average count rate $S_0$ during the acquisition of the data of figure 10, due to all sources (sodium return, Rayleigh return, and dark signal), was 151 msec$^{-1}$.

The in-phase data of figure 10 can be deconvolved and normalized to yield the resonance height $H_{Na}$ and the resonance width $\Delta f$. Figure 11 shows the result of deconvolving the data of figure 10, and normalizing by the amount of DC sodium return signal observed far from resonance. (By DC signal, we mean signal observed directly, rather than using lock-in detection that only can observe an AC signal.) The signal due to Rayleigh and to dark counts were subtracted off, so the data of figure 11 contains only sodium signal, and thus can be compared with the predictions of figures 5 and 6.

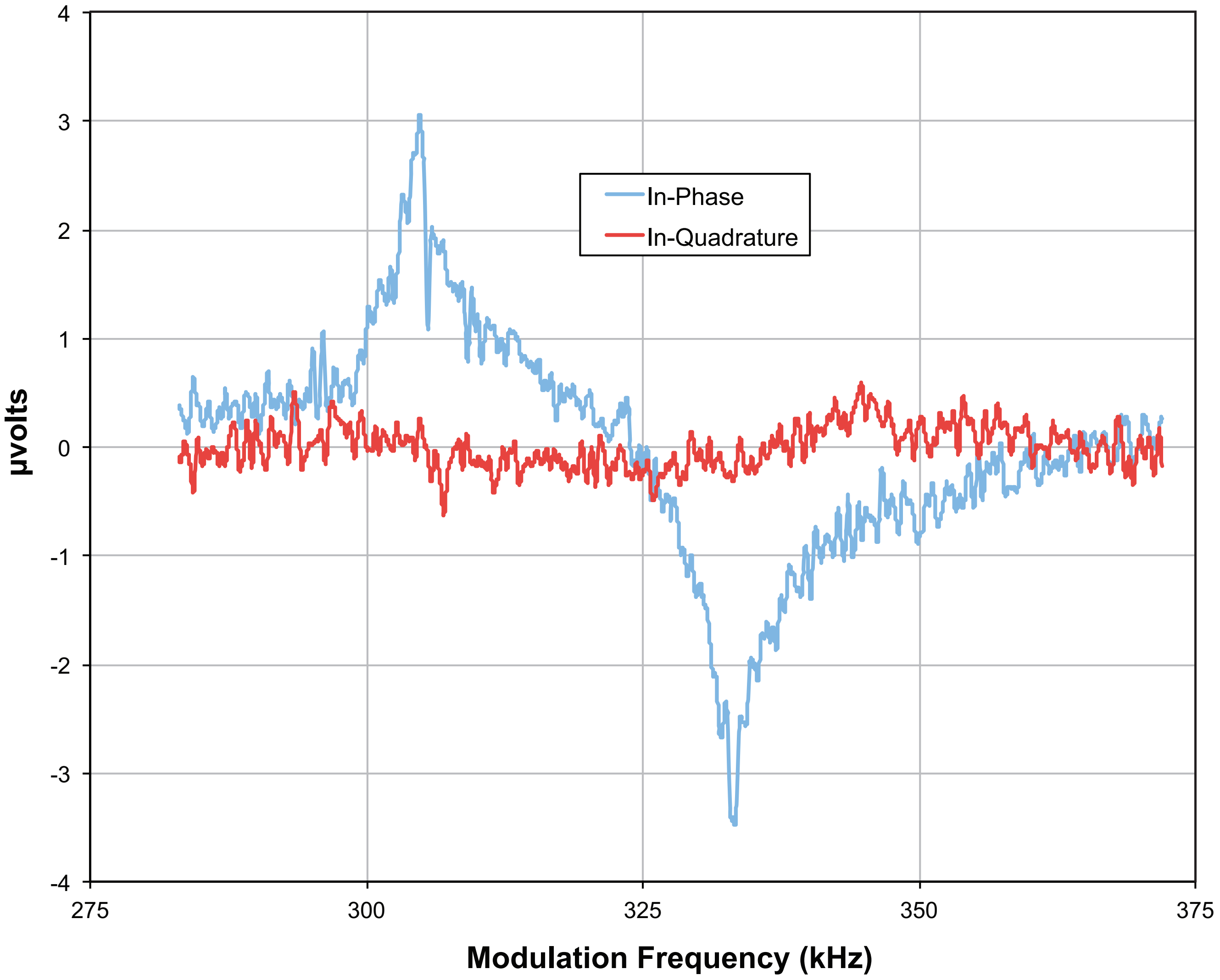

Figure 10. Average PRF was swept from 283 to 372 kHz, while being dithered ±15 kHz. The lock-in detection process creates a positive image of the Larmor resonance translated by -15 kHz, summed with a negative image translated by +15 kHz. Data was taken from 0718 to 0848 UTC on March 25, 2016.

Figure 11 shows that the resonance height, which we have called $H_{Na}$ to remind us that the reference level is the off-resonance signal due to sodium, is about 5.9%. The steepest part of the resonance, which is where you want to operate to be highly sensitive to a magnetic field change, can fit to a triangle function with a FWHM of 4 kHz. (A triangle is a rather simple function that clearly does not fit the data over a wide range of frequencies. But looking back at equation 4, observe that all that really matters for magnetometry is the maximum slope of the resonance. The use of a triangle is a simple way to estimate the peak slope of the data, and to get an effective value of $\Delta f$ that will result in that peak slope.)

The value of $H_{Na}$ is interesting for comparison to the model results of figure 6, but it is not the value of $H$ needed for use in equation 8. For that, we need the change relative to all signal, including that from dark signal and from Rayleigh signal. This yields a value of $H = 2.8\%$. With values of $S_0 = 151\ msec^{-1}$, $H = 2.8\%$ and $\Delta f = 4\ kHz$, and the known value of $g$, we can use equation 8 to calculate an expected system sensitivity of 74 nT/√Hz.

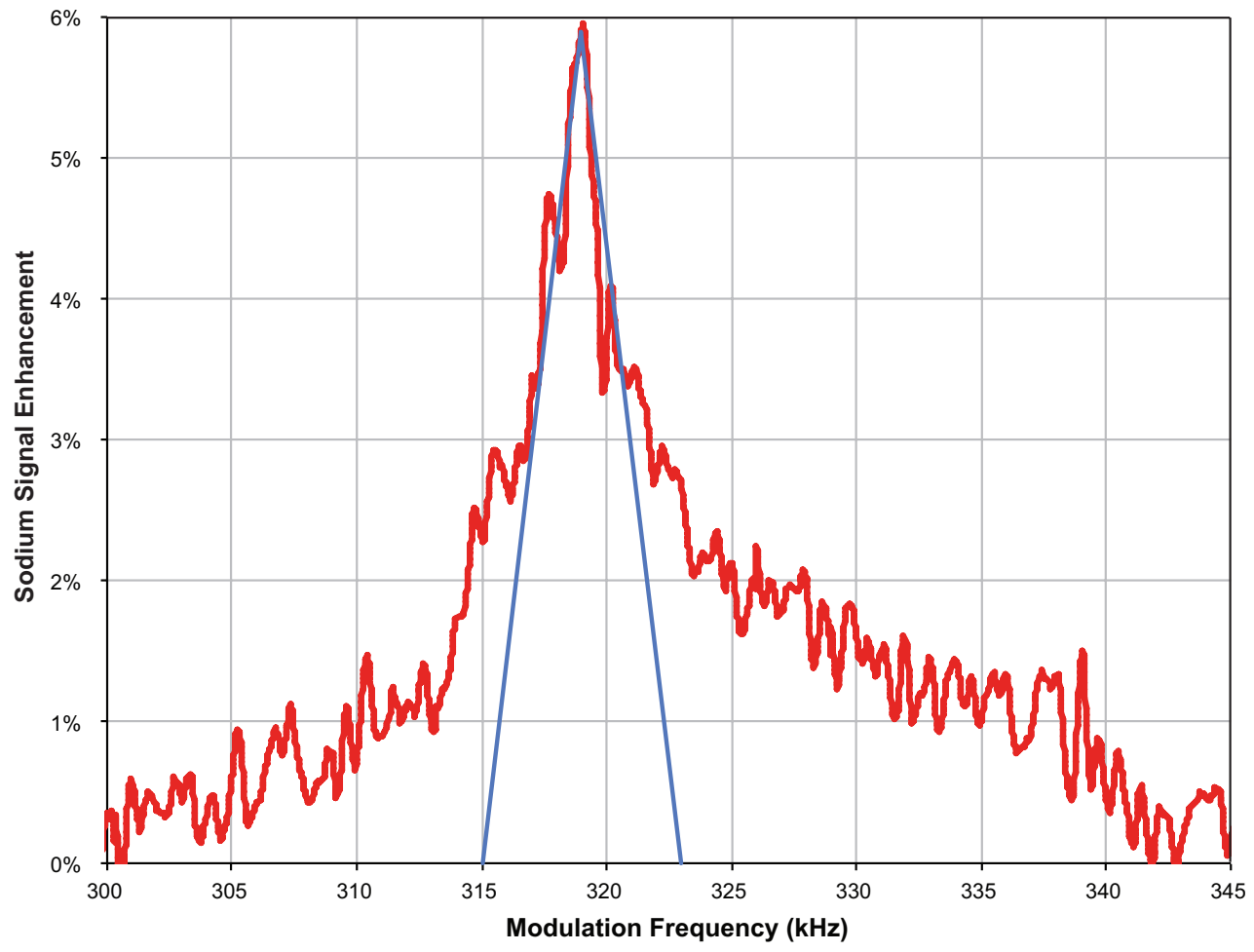

Figure 11. Larmor resonance. Vertical axis is change in the sodium signal relative to the sodium signal far off resonance. Blue curve is a fit to a triangle resonance shape with $H = 5.9\%$ and $\Delta f = 4\ kHz$.

Figure 12 shows how we determined both $S_0$, the average count rate of the detector, and $V_{DC}$, the DC voltage needed for the normalization that led to figure 11. Figure 12 is the direct output of the detector, recorded for a period of 0.2 milliseconds, under conditions identical to those of figure 10. The individual pulses created by detected photons can be seen.

*Magnetic field measurement*

The Larmor resonance peak frequency $f_L$, and thus the magnetic field magnitude $F$, can be estimated from the data of figure 11. Creating figure 11 required a long, slow scan of frequency, with most of the frequencies during the scan being far off resonance, where there is little magnetic signal. If one's goal is to quickly measure small changes in magnetic field, a better approach to the estimation of $f_L$ is to reduce the size of the frequency dither, so that the full amount of frequency dither is $\Delta f$. (That is, a dither of $\pm\Delta f/2$ from the center frequency $f_{avg}$.) If $f_{avg}$ is set at $f_L$, then the lock-in detector will produce a signal of zero, since the signal at $f_L + \Delta f$ and at $f_L - \Delta f$ are the same. (We are assuming a symmetric resonance function. Asymmetry will introduce a slight offset.) Any change in $f_L - f_{avg}$ due to change in magnetic field or due to change in $f_{avg}$ will produce the largest possible change in signal, since both $f_L + \Delta f$ and at $f_L - \Delta f$ are set at the inflection points, where the slope is maximum. This change in signal is exactly that expressed by equation 4.

Figure 13 shows data taken in this manner. The lock-in signal was recorded, for 45 minutes, from 10:11 to 10:56 UTC on March 25, 2016. The value of $f_{avg}$ was set at 318 kHz for most of this data run. The dither value was set to ±2.5 kHz for all of the run, with a dither frequency of 314.159 Hz. From 10:31 until 10:33, $f_{avg}$ was shifted to 316 kHz. From 10:54 until the end of the run at 10:56 $f_{avg}$ was set to 320 kHz.

The two, two-minute long, 2-kHz frequency changes created exactly the signal that would be expected if a magnetic change occurred, with an absolute value of $2\ kHz/g = 286\ nT$. Thus figure 13 provides a direct way to measure the sensitivity of the magnetometer.

The two 2-kHz signal injections, equivalent to 286 nT changes in $F$, created a signal, estimated from figure 13, of 1.75 µV, with the sign of the voltage change the opposite of the sign of the frequency change, and the same as the sign of the "spoofed" magnetic field change. Thus the calibration factor $C$ is $286\ nT / 1.75\ \mu V = 163\ nT/\mu V$. Ignoring the four minutes of injected signal, the rms fluctuation of the data of figure 13 is 0.13 µV, corresponding to 21 nT. The bandwidth of the signal of figure 13 is 0.0167 Hz, as controlled by the settings we applied to the lock-in amplifier. Thus the amplitude spectral density of the background signal of figure 13 is $21\ nT / \sqrt{0.0167\ Hz} = 162\ nT/\sqrt{Hz}$.

We have no reason to believe that any actual fluctuation of magnetic field caused any of the signal in figure 13. March 25, 2016 was magnetically quiet. The field magnitude $F$ as measured by the US Geological Survey (USGS) Tucson Magnetic Observatory (see Acknowledgements) stayed within a range of width 1.02 nT over the 45 minutes of figure 13, small compared to the 21 nT rms noise seen in that figure. Thus it is likely that all of the 162 nT/√Hz background noise is due to the measurement process, and thus that value is a fair characterization of the sensitivity of our instrument. However, it is possible that field fluctuations at 92 km exceed those at ground level, and that some real signal contributes to the "noise."

Figure 14 shows the amplitude spectral density of the background noise of the figure 13 data, that is, the data outside the two 2-minute injected signals. It shows the high variability of non-averaged spectral data, but nothing makes us believe that it is not simply low-passed white noise.

The magnetic field $F$ corresponding to a Larmor frequency $f_L = 318\ kHz$ according to equation 3 is 45441 nT. Ignoring the offset spikes, the average signal during the 45 minutes was -0.005 µV, equivalent to a magnetic field change of -0.8 nT, which is negligible.

*Expected magnetic field*

The USGS operates a magnetic observatory conveniently located for our work. The Tucson Magnetic Observatory is at latitude 32.175° N, 110.734° W, at an elevation of 946 meters. Our magnetometer's laser system transmitter and the Kuiper Telescope receiver were at 32.416° N, 110.735° W, at an elevation of 2510 meters. Our beam was launched with an elevation of 59.7° (90° equals vertical) and an azimuth of 25.1° (east of due north). Table 2 gives the locations where our beam intersected the sodium layer at three altitudes. Also included is the measured or calculated magnetic field from three different models at each location.

The Tucson Magnetic Observatory measured the value of $F = 47404\ nT$ during our 45-minute measurement. The estimated values are from the International Geomagnetic Reference Field, IGRF (Thébault et al., 2015), the World Magnetic Model (NOAA, 2015a), and the SIFMp model (Olsen et al., 2016). The *F* value derived from our measurement, 45441 nT, is consistent with the estimates. Our measured value of $f_L = 318\ kHz$ corresponds to an altitude within but somewhat above the center of the sodium layer. This makes sense in light of the fact that the low pressure of high altitude increases the sodium resonance *H*, as is seen by observing figure 6.

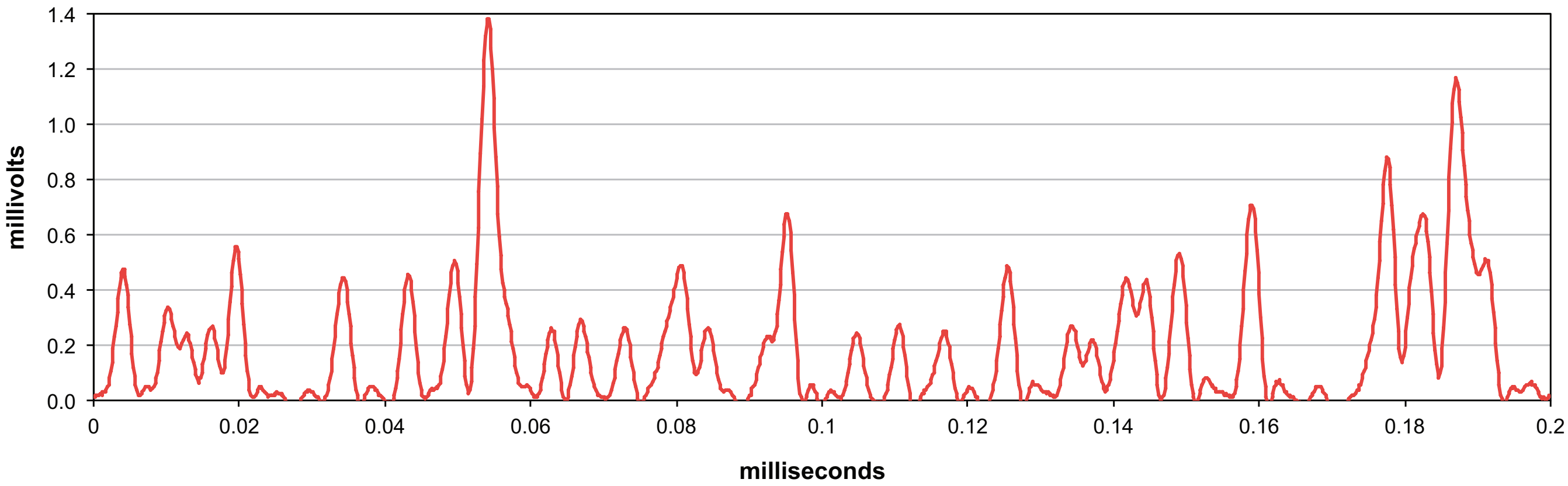

Figure 12. The direct output of the photodetector, under the same conditions as when the data of figure 11 were recorded. Each peak above 0.2 mV is considered as a counted photon. There are 30 counts in 0.2 msec; hence the estimated count rate $S_0$ is 150 msec$^{-1}$. The average value $V_{DC}$ is 0.228 mV.

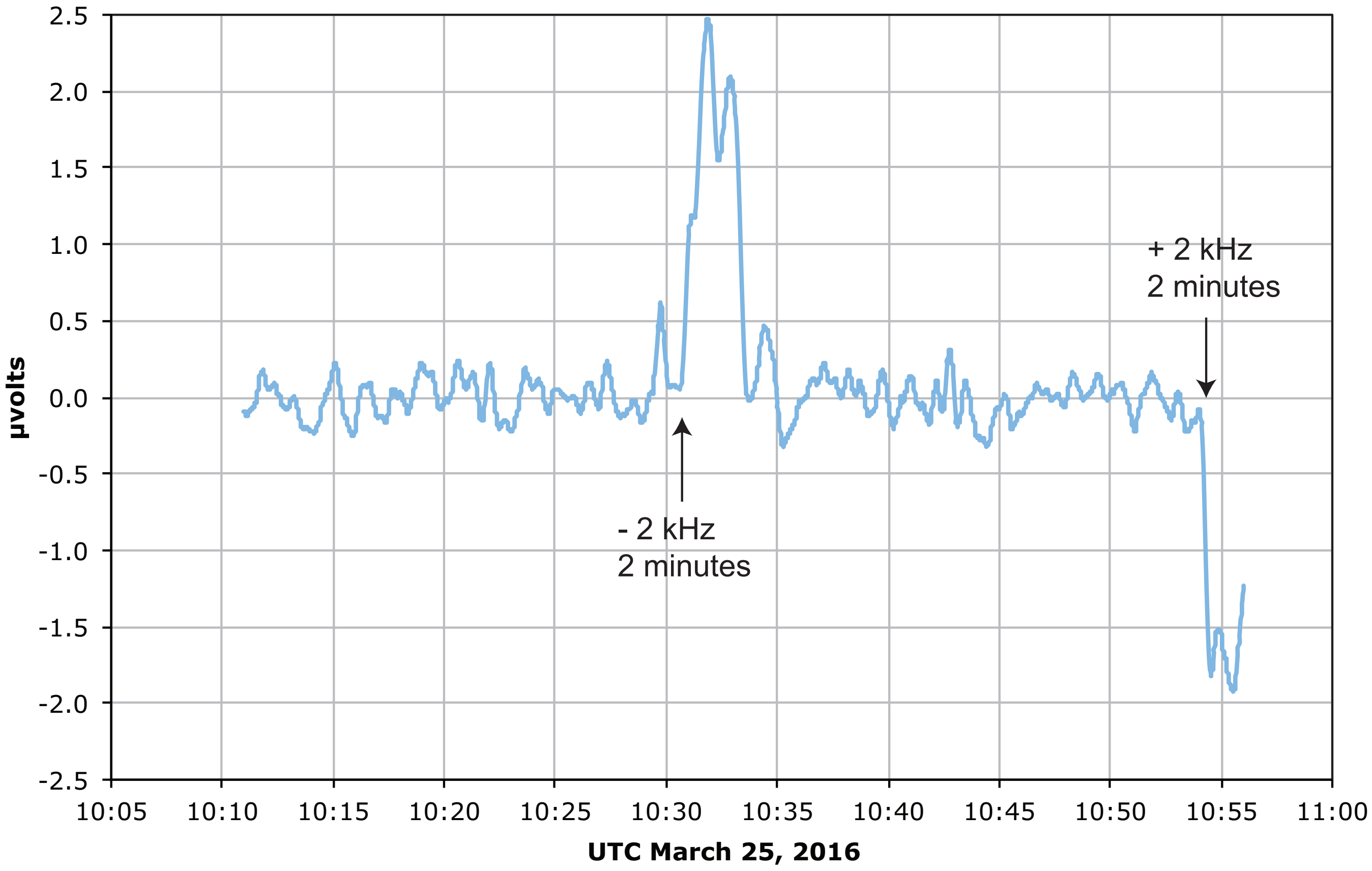

Figure 13. A 45-minute measurement of magnetic field. The PRF was dithered by ±2.5 kHz. For most of the time the center frequency $f_{avg}$ about which the dither occurred was 318 kHz. At the two times indicated by arrows, the center frequency was intentionally offset by -2 kHz or +2 kHz. This "injects" a signal into the system identical to a change in magnetic field of ±286 nT. This allows calibration of the system, and evaluation of the sensitivity.

A model prediction of the magnetic field in the mesosphere is difficult because there are no current observations that can be integrated into a model. Hopefully our described technique and future observations using it can help calibrate these models. All current models use some combination of observations from magnetic observatories at the Earth's surface, and satellite observations from the Swarm Constellation (a European Space Agency mission to study the Earth's magnetic field), currently in polar orbits at 450 and 530 km. This absence of mesospheric observations also makes the calculation of the model uncertainty difficult. The time period of our observation was a very quiet one geomagnetically, with a magnetic disturbance storm index Dst of -2 to -3 nT, and a solar flux activity index F10.7 value of 85.1 s f. u. (NOAA, 2016). This corresponds to expected external fields, and corresponding induced fields, at mesospheric altitudes of about 10 nT (Sabaka et al., 2004). The un-modeled field associated with shorter wavelengths of the crustal field is of order 50 nT in the mesosphere (Lesur et al., 2015). The World Magnetic Model (Chulliat et al., 2015) provides a one-sigma estimate of 152 nT for commission and omission errors.

## 4. Discussion of Results

*Sensitivity*

In previous sections, we have provided three estimates of instrument sensitivity, as measured in nT/√Hz.

The first estimate is based on calculated system parameters. This estimate of 19 nT/√Hz used a value of photon return $S_0$ based on values from table 1, and used values of *H* and $\Delta f$ based on the sodium modeling summarized by figures 5 and 6, with the assumption of optimized beam intensity, and with all signal being sodium signal. Equation 8 then provides the estimate of sensitivity.

The second estimate, 74 nT/√Hz, is based on observed system parameters, and also uses equation 8. It is based on measured values of $S_0$, and values of *H* and $\Delta f$ taken from the fit to the data of figure 11.

The third value, 162 nT/√Hz, was based on actual measurement of both signal and noise, based on the data presented in figure 13.

Table 3 compares the three cases. The tabulated value $H_{Na}$ is the sodium resonance height defined as the height of the resonance divided by the off-resonance background due to sodium scattering. It is related to the resonance height *H* that appears in equation 8 by

$$H_{Na} = H\, S_0 / S_{0,\,Na} \qquad (10)$$

where $S_0$ is the total received signal count rate, including Rayleigh scatter and dark counts, and $S_{0,\,Na}$ is the count rate from sodium scattering alone.

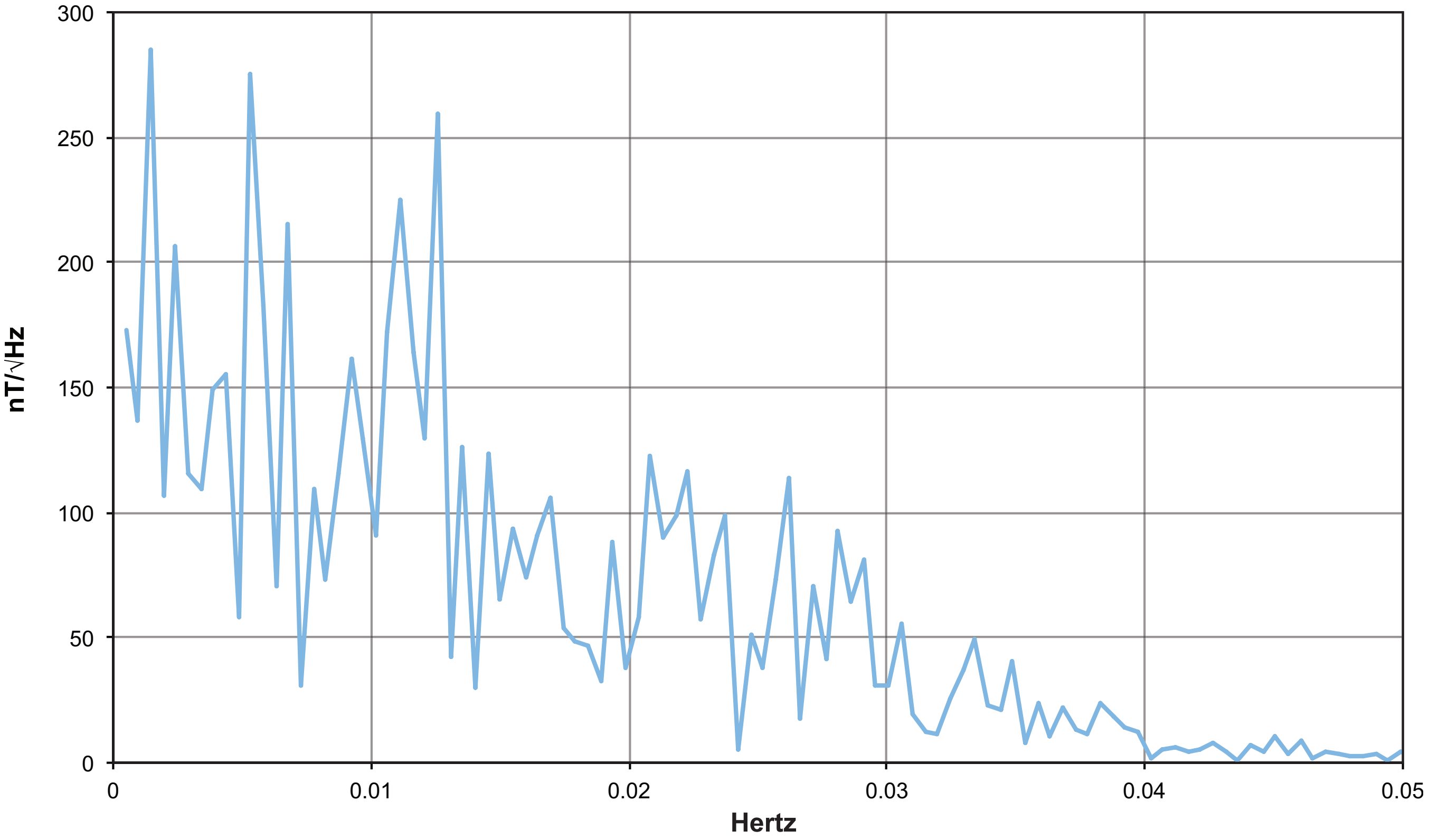

Figure 14. Amplitude spectral density of the data of the previous figure, excluding the injected signals. There is probably no significant magnetic signal, so what is seen is noise, averaging 162 nT/√Hz with a cut-off at 0.0167 hertz. Since this is non-averaged data, the high variability is to be expected.

**Table 2. Locations where beam intersected sodium layer, and ground-based magnetometer reference, plus estimated or measured magnetic field, and resultant $f_L$ based on $F$ using the SIFMp model.**

| | Alt (km) | Lat (°) | Long (°) | $F$ (nT) IGRF | $F$ (nT) WMM | $F$ (nT) SIFMp | $F$ (nT) measured | $f_L$ (kHz) |
|---|---|---|---|---|---|---|---|---|
| Remotely measured value | | | | | | | 45441 | 318.0 |
| Beam crosses bottom of sodium layer | 87 | 32.820 | -110.511 | 45560 | 45551 | 45695 | | 319.8 |
| Beam crosses center of sodium layer | 92 | 32.844 | -110.498 | 45464 | 45456 | 45598 | | 319.1 |
| Beam crosses top of sodium layer | 97 | 32.868 | -110.485 | 45369 | 45361 | 45369 | | 317.5 |
| USGS Tucson Magnetic Observatory | 0.95 | 32.175 | -110.734 | | | | 47404 | 331.7 |

**Table 3. Sensitivity calculated from theoretical and observed system parameters.**

| Case | Total count rate, $S_0$ | Na count rate, $S_{0,\,Na}$ | $H_{Na}$ | $H$ | $\Delta f$ | noise-equivalent signal, using equation 8 |
|---|---|---|---|---|---|---|
| $S_0$, $H$, $\Delta f$ from models | 314 $\text{msec}^{-1}$ | 314 $\text{msec}^{-1}$ | 20% | 20% | 11 kHz | 19 nT/√Hz |
| $S_0$, $H$, $\Delta f$ as measured | 151 $\text{msec}^{-1}$ | 71 $\text{msec}^{-1}$ | 5.9% | 2.8% | 4 kHz | 74 nT/√Hz |
| Effective $S_0$, $H$, $\Delta f$ to fit actual | 102 $\text{msec}^{-1}$ | 48 $\text{msec}^{-1}$ | 5.9% | 2.8% | 7.3 kHz | 165 nT/√Hz |

Our model-based calculation is based on the count rate $S_0$ of 314 $\text{msec}^{-1}$ taken from table 1, with all of $S_0$ assumed to be due to sodium scattering. In fact, as measured, only 71 $\text{msec}^{-1}$ were produced by sodium; 63 $\text{msec}^{-1}$ were due to Rayleigh and 17 $\text{msec}^{-1}$ were dark counts. Since the Rayleigh and dark counts contribute to the background without increasing the resonance, they reduce the useful value of $H$ to 2.8%, according to equation 10.

Thus the discrepancy between the first and second rows of table 3 are due to the following factors:

- At 71 $\text{msec}^{-1}$, the sodium signal was 22% of what we expected.
- We did not expect a significant Rayleigh signal to overlap and interfere with our sodium signal; in fact, it was almost as big as the sodium signal, thus reducing by more than half the effective value of $H$.

- The intensity was probably as low as 0.1 watt/m$^2$ , outside the optimal range of 0.35 watt/m$^2$ to 1 watt/m$^2$. This reduces the value of both $H$ and $\Delta f$, especially at lower altitudes.

We must also discuss the discrepancy between the 74 nT/√Hz sensitivity estimated based on observed system parameters, and the 162 nT/√Hz sensitivity derived directly from the data of figure 13, the two differing by a factor of 2.2. It could be that the noise is higher than expected, or the signal could be lower than expected. (It could also be that we have measured some "real" geomagnetic noise, but we will assume that this is not the case.)

The DC signal $V_{DC}$ observed while the data of figures 10 and 13 were being collected was 228 µV, and this signal was due to a photon count rate $S_0$ of 151 msec$^{-1}$. The detection bandwidth $BW$ was 0.0167 Hz. The theory of Poisson statistics says the expected signal variance $<\Delta V^2>$ based on pure shot noise would be[7]

$$<\Delta V^2> = 2\, V^2_{DC}\, BW / S_0 \qquad (11)$$

which for the parameters given would give $\Delta V = 0.11\ \mu V$. The observed value taken directly from the data of figure 13 was $\Delta V = 0.13\ \mu V$. Thus based on increased noise, the sensitivity would be raised by a factor of $0.13/0.11 = 1.21$.

One way to characterize excess noise in a detector is as a reduced quantum efficiency, leading to a reduced effective count rate for purposes of calculating noise. Looking to equation 8, observe that a reduced value of $S_0$ will increase (that is, degrade) the sensitivity in inverse proportion to $S_0^{1/2}$. Thus the effective value of $S_0$ presented in the bottom row of table 3 is reduced from the value in the row above by a factor of $(1/1.21)^2 = 68\%$. This factor can be thought of as multiplying the quantum efficiency of the detector to create a reduced, effective quantum efficiency of $27\% \times 68\% = 18\%$.

One likely source of this excess noise is the considerable variance in the size of the pulses created by photons in the semiconductor detector, due to the nature of the device (Adamo et al., 2014; Hamamatsu, 2013 and 2016). This can be observed in figure 12. One cause of pulse height variation is double pulsing, caused by the fact that the electron avalanche creates a small amount of light, which can create a second, simultaneous avalanche.

The expected signal can also be estimated once we know $V_{DC}$. The signal $\Delta V$ expected due to an injected frequency change $\Delta f_{injected}$ is given by

$$\Delta V = V_{DC}\, H\, \Delta f_{injected} / \Delta f \qquad (12)$$

where equation 12 results from the simple triangle model of the resonance shape.

We can calculate an expected value of $\Delta V$ based on the parameters extracted from figures 11 and 12. With $V_{DC} = 228\ \mu V$, $H = 2.8\%$, $\Delta f_{injected} = 2\ kHz$, and $\Delta f = 4\ kHz$, equation 11 yields $\Delta V = 3.2\ \mu V$.

When compared to the observed value of 1.75 µV taken from figure 13, the actual signal is lower than expected. This reduction in signal raises (degrades) the sensitivity by a factor of $3.2/1.75=1.83$. Multiplying the increased noise factor of 1.21 by the reduced signal factor of 1.83, we get 2.2, closely matching the observed difference between calculated and observed sensitivity.

[7] *Equation 11 can be derived by starting with equation 5. Then assume that $V_{DC}$ is proportional to $S_0$ and $\Delta V$ is proportional to $\Delta S$, with the same proportionality constant. This yields three equations. Both the proportionality constant and $\Delta S$ can then be algebraically eliminated, and equation 11 will result.*

The reduced signal may be due to the fact that the value of $\Delta f = 4\ kHz$, taken from the triangle fit in figure 11, is optimistic. Such a low value of $\Delta f$ is at the very limit of what our sodium model predicts, even at low intensity and high altitude. A value of $\Delta f = 7.3\ kHz$ would cause equation 12 to predict the observed value of $\Delta V$. This value of $\Delta f$, along with the reduced value of $S_0$ , are used in the last row of table 3, predicting 165 nT/√Hz when plugged into equation 8, close to the observed sensitivity of 162 nT/√Hz.

We believe that the data in general support the theoretical work we have presented, and that our sensitivity is worse than expected primarily because the intensity in the mesosphere was lower than expected, the signal collection efficiency was worse than expected, and the discrimination against Rayleigh scattered light was worse than expected. Also, the noise properties of the detector can be characterized by a reduced quantum efficiency.

All of these issues relate to the optimization of optics – the launch optics, collection optics and detector. In a better optimized and better characterized system, we expect that actual sensitivity would much more closely match theory.

*Improving sensitivity and resolution*

The sensitivity, measured in nT/√Hz, is important for improving the system's ability to measure a time-varying field. A system with a sensitivity of 1 nT/√Hz would provide a sensitivity approaching that of typical observatory-grade magnetometers, but at a previously inaccessible altitude.

There are many ways in which our system could be improved. Two of the clearest are an improvement in the quality of the launched beam, so that we could be working at the optimum beam intensity, and an improvement of the efficiency of the collection optics. In both respects, our results are well below the state of the art, due to time and cost constraints, and it should be possible to make improvements.

In order to optimize the launch optics, we would need to interferometrically analyze the launch telescope, and find if the wavefront were close enough to flat to create a diffraction-limited beam. Then we would replace components shown to be inadequate. In order to improve the efficiency of the collection optics, we would need to improve the throughput of each component in the receive telescope. We would also add an aperture or other mechanism to exclude Rayleigh light. We would begin by measuring total throughput, by observing the signal from a star of known magnitude. All these improvements were attempted, but we were unable to get useful results due to weather delays and within the time and cost limits of this project.

Another improvement would be to use a state-of-the-art guidestar laser. Such lasers, which are commercially available, have a power of 20 watts, and are optimized for high photon return. (An important contributor to increased photon return is the addition of a laser wavelength, about 10% in power, resonant with the $D_{2b}$ line of sodium. This wavelength "repumps" the sodium and increases $\Psi$ to a value of 450 photons/second/steradian/atom/(W/m$^2$) substantially above the value of 240 as shown in table 1, which is valid for our single-wavelength laser. Holzlöhner et al. (2010) includes theory and calculations of the effect of sodium atom repump.) The expected return from such lasers is 36 x 10$^6$ photons/second/m$^2$ of incident light at the telescope aperture when the laser is operating continuously at its rated power of 20 watts (Drummond et al., 2007). These lasers are becoming more common at the large, ad-vanced astronomical observatories, which have collection areas of 50 $m^2$ or more.

A further improvement in sensitivity is theoretically possible if the laser used has a broadened linewidth. A laser with a linewidth roughly equal to the Doppler-broadened sodium linewidth, that is, about 500 MHz, will interact with all of the sodium atoms, instead of only one velocity class. For a narrow linewidth laser, any velocity-changing collision by a sodium atom results in the atom being lost from the set of precessing atoms interacting with the laser beam, even if the angular momentum of the atom is unchanged by the collision. The linewidth $\Delta f$ of the Larmor resonance is proportional to the rate at which atoms are lost from this set. A broad linewidth laser continues to interact with an atom after a change in its velocity, if the angular momentum state is unchanged.

It turns out that collisions with $O_2$ change angular momentum, but collisions with $N_2$ do not (Milonni et al., 1999; Holzlöhner et al., 2010). The atmosphere is about 20% $O_2$. Thus even with a broad laser linewidth, about 20% of all collisions will remove the atom from the set that is precessing in phase with the laser. The Larmor linewidth assuming that only $O_2$ collisions contribute is expected to be about 20% of what it would be if all collisions contributed to linewidth.

A pulsed laser with broadened linewidth has been proposed and patented (Kane, 2014, Kane et al., 2015). The proposed design has the additional advantage of greater power in a pulsed mode. A disadvantage of the broadened linewidth is that the optimum mesospheric intensity is increased, meaning that the beam has to be smaller in the mesosphere or the laser increased in power.

Table 4 summarizes three design options for remote magnetometers with improved sensitivity. All assume improved launch beam quality and collection efficiency, and assume that the beam intensity in the mesosphere is optimized. A detector with 0.27 counts per incident photon noise-effective quantum efficiency is assumed, such as a photomultiplier tube or an improved MPPC. An optimized 20-watt guidestar system with $\Psi$ = *450 photons/second/steradian/atom/(W/m²)* is assumed throughout table 4, other values are as from table 1.

**Table 4. Projected sensitivity of improved remote laser magnetometers**.

| System | Current with improved optics and laser | Case 1 with broadened linewidth laser | Case 1 with large receive telescope |
|---|---|---|---|
| Telescope aperture | 1.8 $m^2$ | 1.8 $m^2$ | 50 $m^2$ |
| Laser average power | 7 watts | 20 watts | 7 watts |
| Laser linewidth | <1 MHz + $D_{2b}$ | 500 MHz + $D_{2b}$ | <1 MHz + $D_{2b}$ |
| Detected signal, effective | 3095 counts/msec | 8842 counts/msec | 85964 counts/msec |
| Resonance height *H* | 20% | 20% | 20% |
| Resonance width $\Delta f$ | 11 kHz | 2.2 kHz | 11 kHz |
| Sensitivity using equation 8 | 6.3 nT/√Hz | 0.7 nT/√Hz | 1.2 nT/√Hz |

The first option would make use of a state-of-the art cw single-frequency laser, with a telescope similar to what we used. Laser power is 35% of 20 watts, because the average power launched from the 20-watt laser is reduced by the 35% duty cycle of the pulse format. The second option assumes that the laser is a broad linewidth, naturally-pulsed design, so that no duty cycle penalty is incurred. The third option assumes that the state-of-the-art cw laser is used with a large telescope, the type that is typically equipped with a guidestar laser.

Fan et al. (2016) have made an interesting and theoretically sound proposal for avoiding the duty-cycle penalty. Instead of pulsing the laser near the Larmor frequency, they would reverse the sense of the circular polarization at that frequency. The only drawback is the need for fast polarization modulation of the high powered 589-nm beam.

*Spatial Resolution*

According to table 2, the value of $f_L$ changes by 1.3 kHz along the path of the laser beam through the layer. The sodium atoms at lower altitude have a higher value of $f_L$, and a larger linewidth $\Delta f$.

It would be possible to get range-resolved measurement of magnetic field. This could be done by creating a greater separation between the transmit telescope and the receive telescope, so that the receive telescope sees the transmitted beam passing through its field of view at a slight angle. The simple geometry needed to understand this is shown in figure 15. If the desired apparent size of the beam in the sodium layer is *x*, and the distance to the sodium layer is *Z*, and the thickness of the sodium layer is *L*, then the needed separation *S* of the transmit and receive telescopes is given by

$$S = x\,Z\,/\,L. \qquad (13)$$

For *Z* = *106 km, L* = *10 km, and x* = *10 meters, S* = *106 meters*. With our current spot size of 2.6 meters, these values would lead to four distinct resolvable spots where the magnetic field could be measured independently. Of course, the return signal from each spot would be only one-quarter of the total signal, increasing sensitivity by a factor of two at each range.

Another way to get altitude resolved measurements would be to lower the PRF to a subharmonic of the Larmor frequency. The spin relaxation time (250 µs) is much longer than 10x the Larmor period (32 µs). A PRF of 31.8 kHz would produce only one pulse in the sodium layer at a time and the altitude detected could be determined by time-gating the detector.

Either of the two methods could eliminate the undesirable Rayleigh backscatter photon flux.

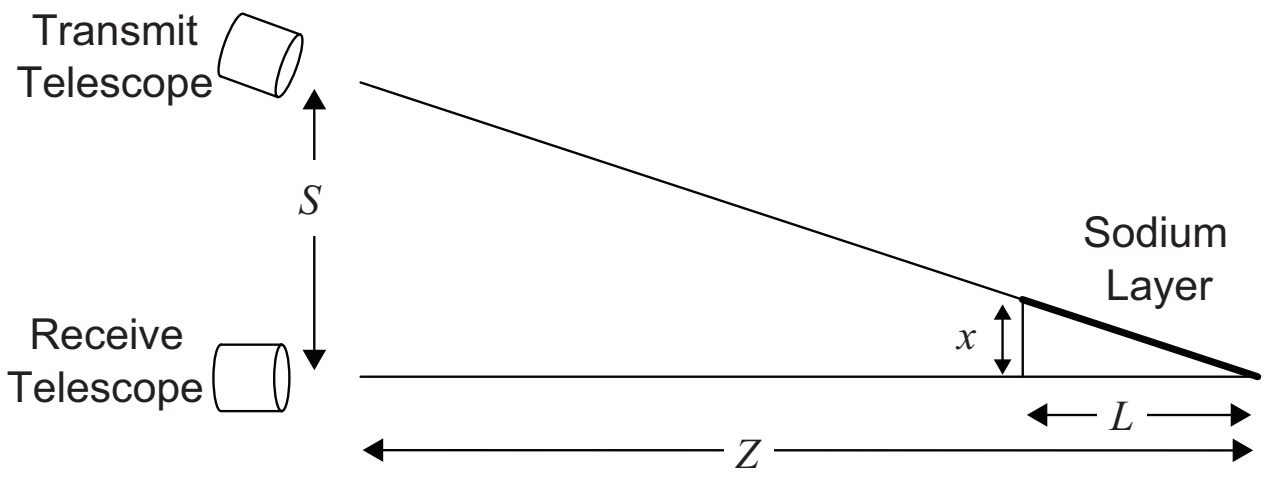

Figure 15. Geometry enabling range-resolved magnetometry, by placing transmit telescope distant from receive telescope, creating an extended source in the view of the receive telescope. The value *x* is the apparent size in meters of the beam in the sodium layer, as seen from the receive telescope.

## CONCLUSIONS

We have demonstrated a remote magnetometer that has measured the magnetic field in the Earth's mesosphere, using sodium atoms at a 106-kilometer distance from our instrument. A pulsed laser with average transmitted power of 1.33 watts illuminated the sodium atoms, and the field was inferred from back-scattered light collected by a telescope with a 1.55-meter-diameter aperture. The magnetic field was measured with a sensitivity characterized by an equivalent noise spectral density of 162 nT/√Hz. At this sensitivity, a measurement based on one hour of data averaging would have an expected rms error due to noise of 2.7 nT. The value of magnetic field inferred from our measurement, 45441 nT, is consistent with a value estimated based on the Earth's modeled field shape to within a fraction of a percent.

Projected improvements in optics could lead to sensitivity of 20 nT/√Hz, and the use of advanced lasers or a large telescope could approach 1-nT/√Hz sensitivity. All experimental and theoretical sensitivity values are based on a 60° angle between the laser beam axis and the magnetic field vector; at the optimal 90° angle sensitivity would be improved by about a factor of two. But we were limited to a 60° angle at the Kuiper Telescope, so we consistently use that value. The optimal 90° angle may only be possible in the tropics; 60° or better should be possible over most mid-latitude locations.

This technique works by tracking a resonance in the return signal that appears when the sodium atoms are illuminated by a circularly-polarized laser pulsing at the frequency of the sodium atom precession, typically in the range 200 - 350 kHz. For the single-frequency laser we used, the height and width of the resonance reaches an optimum when the average intensity of the pulsed laser light at the atoms is about 0.5 W/m$^2$. Our system was below this optimum, closer to 0.1 W/m$^2$.

Because neither satellites nor balloons can reach this altitude, this technique creates a unique window into an interesting zone of the geomagnetic field, on the lower edge of the ionosphere. Only rockets have been able to reach this zone previously, and have only made transient measurements, not sustained measurements needed to yield the geomagnetic background noise (Sesiano and Cloutier, 1976; Volland, 1984).

Sodium layers exist in other planets and moons in our solar system. It should be possible to use this technique from a satellite in orbit, provided that the orbit is low enough to create a reasonably small beam in the sodium layer without the use of large optics. The fact that this technique is ultimately a measurement of frequency means that very long averaging times can be used, since frequency references can be extraordinarily stable. Thus it is possible to imagine useful data resulting even at extremely low levels of return signal or during the daytime with higher background.

## ACKNOWLEDGMENTS

This material is based upon work supported by the U.S. Navy, Office of Naval Research - Code 321 under contract numbers N00014-11-C-0314 and N00014-14-C-0110. The Final Report for the earlier contract provided estimates for both the sensitivity of the laser-based remote measurement under various conditions, and for the correlation that would be observed between high-altitude data and surface data.

The results presented in this paper rely on data collected at the Tucson Magnetic Observatory. We thank the US Geological Survey for supporting its operation and INTERMAGNET for promoting high standards of magnetic observatory practice (www.intermagnet.org), and; the National Geomagnetism Program (observatory location Tucson, AZ):
http://geomag.usgs.gov/monitoring/observatories/tucson/
http://geomag.usgs.gov/map/#realtime
http://geomag.usgs.gov/plots/

We'd also like to acknowledge the University of Arizona, Steward Observatory, for use of the Kuiper Telescope and the 589-nm FASOR coherent light source. Special thanks go to Jim Grantham and Steve Bland of the Stewart Observatory Mount Operations Team for making some last minute unscheduled telescope configuration changes, and finally Duncan Reed for weathering several cold nights airplane spotting.